\documentclass[sigconf,natbib=true,screen=true]{acmart}

\acmSubmissionID{6661}

\usepackage{color}
\usepackage{bbm}
\usepackage{multirow}
\usepackage[inline]{enumitem}
\usepackage{graphicx}
\usepackage{subfigure} 
\usepackage{sistyle}
\usepackage[ruled]{algorithm2e}
\usepackage{booktabs}
\usepackage{caption}
\SIthousandsep{,}
\usepackage{makecell}
\usepackage[skip=0pt]{caption}
\usepackage{amsmath}
\usepackage{hyperref}
\usepackage{array} 
\usepackage{graphicx}
\usepackage{acronym}
\usepackage[export]{adjustbox}

\newcommand{\heading}[1]{\vspace*{0.5mm}\noindent\textbf{#1.}}
\AtBeginDocument{%
  \providecommand\BibTeX{{%
    \normalfont B\kern-0.5em{\scshape i\kern-0.25em b}\kern-0.8em\TeX}}}

\makeatletter
\g@addto@macro\normalsize{%
  \abovedisplayskip 2pt plus1pt 
  \belowdisplayskip 2pt plus1pt
  \abovedisplayshortskip  2pt plus1pt%
  \belowdisplayshortskip  1pt plus1pt
}
\makeatother

\acrodef{IR}{infor\-mation retrieval}
\acrodef{LLM}{large language model}
\acrodef{QA}{question answering}

\copyrightyear{2024}
\acmYear{2024}
\setcopyright{rightsretained}
\acmConference[SIGIR '24]{Proceedings of the 47th International ACM SIGIR Conference on Research and Development in Information Retrieval}{July 14--18, 2024}{Washington, DC, USA}
\acmBooktitle{Proceedings of the 47th International ACM SIGIR Conference on Research and Development in Information Retrieval (SIGIR '24), July 14--18, 2024, Washington, DC, USA}
\acmDOI{10.1145/3626772.3657784}
\acmISBN{979-8-4007-0431-4/24/07}
\author{Hengran Zhang}
\orcid{0009-0004-1144-1298}
\affiliation{
	\institution{CAS Key Lab of Network Data Science and Technology, ICT, CAS}
	\institution{University of Chinese Academy of Sciences}
	\city{Beijing}
	\country{China}
}
\email{zhanghengran22z@ict.ac.cn}

\author{Ruqing Zhang}
\orcid{0000-0003-4294-2541}
\affiliation{
	\institution{CAS Key Lab of Network Data Science and Technology, ICT, CAS}
	\institution{University of Chinese Academy of Sciences}
	\city{Beijing}
	\country{China}
}
\email{zhangruqing@ict.ac.cn}

\author{Jiafeng Guo}
\orcid{0000-0002-9509-8674}
\authornote{Jiafeng Guo is the corresponding author.}
\affiliation{
	\institution{CAS Key Lab of Network Data Science and Technology, ICT, CAS}
	\institution{University of Chinese Academy of Sciences}
	\city{Beijing}
	\country{China}
}
\email{guojiafeng@ict.ac.cn}

\author{Maarten de Rijke}
\orcid{0000-0002-1086-0202}
\affiliation{
 \institution{University of Amsterdam}
 \city{Amsterdam}
 \country{The Netherlands}
}
\email{m.derijke@uva.nl}

\author{Yixing Fan}
\orcid{0000-0003-4317-2702}
\affiliation{
	\institution{CAS Key Lab of Network Data Science and Technology, ICT, CAS}
 \institution{University of Chinese Academy of Sciences}
 \city{Beijing}
 \country{China}
}
\email{fanyixing@ict.ac.cn}

\author{Xueqi Cheng}
\orcid{0000-0002-5201-8195}
\affiliation{
	\institution{CAS Key Lab of Network Data Science and Technology, ICT, CAS}
	\institution{University of Chinese Academy of Sciences}
	\city{Beijing}
	\country{China}
}
\email{cxq@ict.ac.cn}

\renewcommand{\shortauthors}{Hengran Zhang et al.}
\begin{document}

\title[Are Large Language Models Good at Utility Judgments?]{Are Large Language Models Good at Utility Judgments?}

\begin{abstract} 
Retrieval-augmented generation (RAG) is considered to be a promising approach to alleviate the hallucination issue of large language models (LLMs), and it has received widespread attention from researchers recently. Due to the limitation in the semantic understanding of retrieval models, the success of RAG heavily lies on the ability of LLMs to identify passages with utility. Recent efforts have explored the ability of LLMs to assess the relevance of passages in retrieval, but there has been limited work on evaluating the utility of passages in supporting question answering.

In this work, we conduct a comprehensive study about the capabilities of \acp{LLM} in utility evaluation for open-domain \ac{QA}.
Specifically, we introduce a benchmarking procedure and collection of candidate passages with different characteristics, facilitating a series of experiments with five representative LLMs. 
Our experiments reveal that: 
\begin{enumerate*}[label=(\roman*)]
\item well-instructed \acp{LLM} can distinguish  between relevance and utility, and that \acp{LLM} are highly receptive to newly generated counterfactual passages.
Moreover, 
\item we scrutinize key factors that affect utility judgments in the instruction design. 
And finally, 
\item  to verify the efficacy of utility judgments in practical retrieval augmentation applications, we delve into \acp{LLM}' QA capabilities using the evidence judged with utility and direct dense retrieval results. 
\item We propose a $k$-sampling, listwise approach to reduce the dependency of \acp{LLM} on the sequence of input passages, thereby facilitating subsequent answer generation. 
We believe that the way we formalize and study the problem along with our findings contributes to a critical assessment of retrieval-augmented \acp{LLM}. Our code and benchmark can be found at \url{https://github.com/ict-bigdatalab/utility_judgments}.
\end{enumerate*}
\end{abstract}

\vspace{-2mm}

\maketitle

\acresetall

\section{Introduction}
Retrieval-augmented generation (RAG) is considered a crucial means to effectively mitigate the hallucination issues in \acp{LLM} \cite{shuster2021retrieval, ram2023context, ren2023investigating, zamani2022retrieval, xie2023adaptive}, garnering widespread attention from researchers recently \cite{izacard2022few, shi2023replug, lewis2020retrieval, yu2022generate}. 
Due to the semantic limitations in the retrieval model's understanding, the practical efficacy of RAG heavily relies on the \acp{LLM}'s ability to accurately identify passages with utility among the retrieved candidate passages. 
Recent studies have explored the capabilities of \acp{LLM} in relevance judgments during retrieval \cite{xie2023factual, faggioli2023perspectives, pradeep2023rankvicuna, sun2023chatgpt, zhuang2023setwise, pradeep2023rankvicuna}. 
However, there has been limited focus on evaluating the utility judgment.

\begin{figure}[t]
    \centering
    \includegraphics[width=\linewidth]{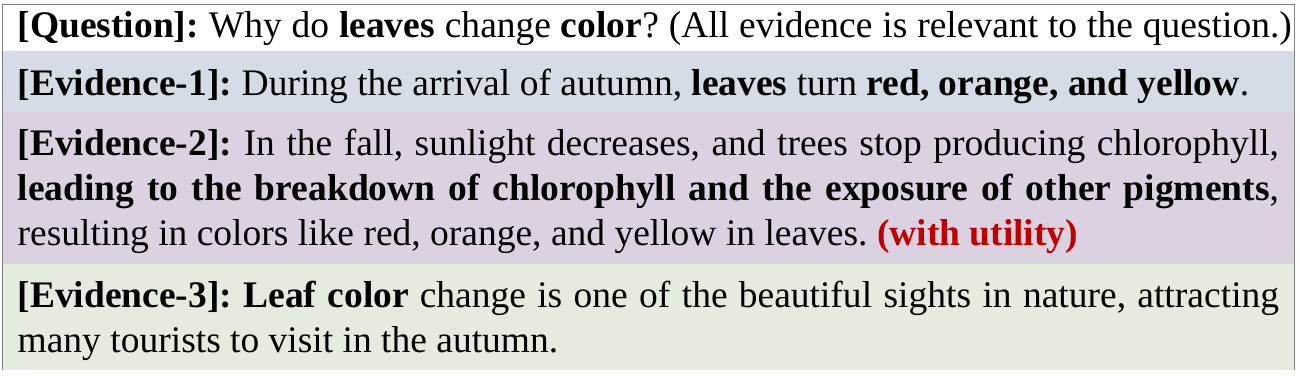}
    \vspace{-2mm}
    \caption{An example between utility and relevance.}
    \label{fig:intro_case}
\end{figure}

\heading{Relevance judgments via \acp{LLM}} 
\citet{faggioli2023perspectives} investigated the use of \acp{LLM} to automatically generate relevance judgments in \ac{IR}. 
Many works have examined the zero-shot language understanding and reasoning capabilities of \acp{LLM} for relevance ranking, including pointwise \cite{zhuang2023beyond, zhuang2023beyond}, pairwise \cite{qin2023large, jiang2023llm}, and listwise \cite{ pradeep2023rankvicuna, sun2023chatgpt, zhuang2023setwise} approaches. 
These publications have shown that \acp{LLM} excel in judging the relevance of passages to the query and achieve state-of-the-art ranking performance on standard benchmarks \cite{zhuang2023setwise, sun2023chatgpt, qin2023large}.

\heading{Utility judgments via \acp{LLM}} 
In this paper, we explore whether \acp{LLM} are good at judging the utility of passages. 
To appreciate the importance of (passage) utility, recall that \acp{LLM} find extensive use in open-domain \ac{QA} \cite{ren2023investigating, yu2022generate, izacard2022few, shi2023replug}. 
A common approach in open-domain QA involves retrieval-augmented \acp{LLM}~\cite{zhang2023hybrid, lewis2020retrieval, xie2023factual}, which 
\begin{enumerate*}[label=(\roman*)]
    \item acquires a set of supporting evidence upon which to condition, and
    \item incorporates the selected evidence into the downstream \ac{LLM} generation process. 
\end{enumerate*}
In open-domain \ac{QA}, the goal of obtaining supporting evidence differs significantly from the goal of obtaining relevance judgments using \acp{LLM}: the supporting evidence should align with the judgments made by the \ac{LLM} regarding which evidence qualifies as having \emph{utility} for answering the question \cite{zhang2023relevance}.  
Utility and relevance are distinct concepts \cite{liu2023webglm}:
\begin{enumerate*}[label=(\roman*)]
    \item \emph{Relevance} signifies a connection between information and a context or question~\citep{saracevic2016notion}, and 
    \item \emph{utility} refers to the practical benefits of downstream tasks derived from consuming the information \cite{saracevic1988study}. 
    E.g.,  all evidence in Fig. \ref{fig:intro_case} has different relevance to the question. 
    However, only ``[Evidence-2]'' has utility in answering the question and other evidence, although connected to ``leaf color change'', lacks useful information on the underlying reasons for this phenomenon.
\end{enumerate*}

\heading{Research goal}
The above observations naturally raise a question: \emph{Are \acp{LLM} not only good at generating relevance judgments but also utility judgments?}
To address this question, we undertake an empirical study in the setting of open-domain QA, investigating the capability of different \acp{LLM} in judging passage utility. Specifically, our study is broken down into three concrete research questions:
\begin{enumerate*}[leftmargin=*,label=(\textbf{RQ\arabic*}),nosep]
    \item \emph{Can \acp{LLM} distinguish between utility and relevance?}

    \item \emph{What factors affect the ability of \acp{LLM} to judge  the utility of evidence?}
    
    \item \emph{How do utility judgments impact the QA abilities of retrieval-augmented \acp{LLM}?}    
\end{enumerate*}

\heading{Benchmarking procedure} 
In this work, we introduce the utility judgments task: 
\emph{Given a question and a set of candidate passages, the utility judgments task is to identify supporting evidence with utility in answering the question}. 
We use several representative \acp{LLM} as zero-shot utility judges.
As illustrated in Figure~\ref{fig:prompt}, we carefully design pointwise, pairwise, and listwise prompting approaches for \ac{LLM}-based utility judges, as well as QA prompting approaches guiding \ac{LLM} in answering questions using selected evidence.

To facilitate the study and evaluation of the utility judgments task, we formulate two hypotheses guiding the construction of novel benchmark datasets in Section \ref{benchmark_construction}: 
\begin{enumerate*}[label=(\roman*)]
    \item \emph{ground-truth inclusion}: the set of candidate passages must encompass ground-truth evidence. The ground-truth evidence offers the highest utility for question among the candidate passages; and 
    \item \emph{ground-truth uncertainty}: there exists uncertainty regarding the presence of ground-truth evidence in the set of candidate passages. 
\end{enumerate*}

Our empirical work leads to the following interesting results: 
\begin{itemize}[leftmargin=*,itemsep=0pt,topsep=0pt,parsep=0pt]
    \item For \textbf{RQ1}: The answer is YES. 
    \acp{LLM} can distinguish between utility and relevance given candidate passages. 
    Specifically, utility judgments may offer more valuable guidance than relevance judgments to \acp{LLM} in identifying ground-truth  evidence necessary for answering questions. 
    Moreover, \acp{LLM} may exhibit a preference for selecting ground-truth evidence with utility when confronted with entity substitution-based counterfactual passages, compared to generated counterfactual passages.

    \item For \textbf{RQ2}: Different \acp{LLM} exhibit varying capabilities in utility judgment, with ChatGPT standing out as the most powerful. 
    There is a consistent improvement in utility judgments performance the expansion of model scale. 
    Listwises approaches may demonstrate superior performance compared to pointwise and pairwise approaches.
    In listwise approaches, \acp{LLM} are sensitive to the position of the ground-truth evidence in the input list. 
    Moreover, the inclusion of chain-of-thought, reasoning process and answer generation also impact performance.

    \item For \textbf{RQ3}: Employing \acp{LLM} as zero-shot utility judges or relevance judges proves more advantageous for answer generation than directly using dense retrieval. 
    The \ac{QA} performance of \acp{LLM} is optimal when using evidence with utility judged by \acp{LLM}.  
    To reduce the dependency of \acp{LLM} on the position of ground-truth evidence, we propose a $k$-sampling listwise approach that combines multiple utility judgments results to derive the final  outcome, thereby facilitating subsequent answer generation. However, a significant gap still exists when compared to using only ground-truth evidence. 
    
\end{itemize}
\noindent%
In Section \ref{problem_statement}, we provide a detailed description of the analysis setting. In Section \ref{exp-1}--\ref{exp-3}, we address the three proposed research questions based on respective experimental results. Section \ref{related_work} discusses related work and conclusions are drawn in Section \ref{conclusion}.

\vspace{-2mm}
\section{Problem statement} \label{problem_statement}

\vspace{-1mm}
\subsection{Task description} \label{task_description}

We introduce the utility judgments task, designed to assess the capabilities of \acp{LLM} to select supporting evidence with utility, which is useful for downstream answer generation. 
Formally, given a question $q$ and a set of $N$ retrieved passages $\mathcal{D}=\{d_1, d_2, \ldots, d_N\}$, the \emph{utility judgments task} is to identify a subset $\mathcal{D}_u$ of $\mathcal{D}$ with passages with utility by prompting \acp{LLM}. We explore two evaluation scenarios based on the assumptions about $\mathcal{D}$: 
\begin{itemize}[leftmargin=*,itemsep=0pt,topsep=0pt,parsep=0pt]

    \item \textbf{Ground-truth inclusion (GTI)}: $\mathcal{D}$ should include ground-truth supporting evidence and other non-ground-truth passages. This assumption facilitates a direct assessment of the accuracy of selected evidence by \acp{LLM} with utility.  

    \item \textbf{Ground-truth uncertainty (GTU)}: Taking into account the practical application of retrieval-augmented \acp{LLM}, $\mathcal{D}$ is directly obtained from the passage retriever without certainty regarding the presence of ground-truth evidence. Consequently, we evaluate the performance of \acp{LLM} in answering questions based on the selected evidence.

\end{itemize}

\vspace{-2mm}
\subsection{Benchmark construction} \label{benchmark_construction}
We introduce the source datasets and retrievers, and outline the construction processes for GTI and GTU.

\vspace{-2mm}
\subsubsection{Source datasets}
\label{source_datasets}
We use three factoid QA (\textbf{FQA}) and non-factoid QA  (\textbf{NFQA}) datasets as our source datasets. 


    In FQA datasets, answers are typically brief and concise facts, such as named entities \cite{hashemi2020antique}: 
    \begin{enumerate*}[label=(\roman*)]
        \item \textbf{Natural Questions (NQ)} \cite{kwiatkowski2019natural} consists of real questions issued to the Google search engine. Each question comes with an accompanied Wikipedia page with an annotated long answer (a paragraph) and a short answer (one or more entities). The long answers are denoted as  ground-truth evidence, while the short answers are denoted as correct answers.
        
        \item \textbf{HotpotQA} \cite{yang2018hotpotqa} consists of QA pairs requiring multi-hop reasoning gathered via Amazon Mechanical Turk, each accompanied by a set of supporting evidence as ground-truth evidence.
        \end{enumerate*}
        
        In NFQA datasets, the answer to a non-factoid question is a chunk of one or more adjacent words \cite{aghaebrahimian2018linguistically}: 
         \textbf{MSMARCO-QA} \cite{nguyen2016ms} is generated by sampling queries from Bing's search logs, consisting of annotated evidence that contains useful information for answering the questions and natural language answers. We use the evidence that is labeled as $is\_selected$: 1 as our ground-truth. 
Following \cite{ren2023investigating, wu2022neural}, our experiments are conducted on the test set of NQ, and the development set of MSMARCO-QA and HotpotQA.

\vspace{-1mm}
\subsubsection{Retriever}
We use two representative dense retrievers to gather supporting passages for the subsequent construction process of benchmark datasets. 
Specifically, we employ RocketQAv2 \cite{ren2021rocketqav2} and ADORE \cite{zhan2021optimizing} for the FQA datasets (NQ and HotpotQA) and the NFQA dataset (MSMARCO-QA), respectively. 
Like \cite{ren2023investigating, shi2023replug, yu2022generate}, we assume that the size of $\mathcal{D}$, i.e., $N$, is 10, which falls within the input scope of the \acp{LLM}. 
In the following, we detail how to build $\mathcal{D}$, for GTI and GTU, respectively.

\vspace{-1mm}
\subsubsection{GTU benchmark}
For three source datasets, we directly use the top 10 passages from the retrieved results for each question as the supporting evidence $\mathcal{D}$.

\vspace{-1mm}
\subsubsection{GTI benchmark}
\label{benchmark}
For three source datasets, the ground-truth evidence is introduced in Section \ref{source_datasets}. 
For non-ground-truth passages, we consider different characteristics as follows. 

\heading{Counterfactual passages (CP)}
We construct synthetic passages that incorporate counter-answers,  conflicting with correct answers. 
In order to make passages contain factual errors, for NQ and HotpotQA, we directly substitute the correct entities in the ground-truth evidence; for MSMARCO-QA, we instruct a LLM to directly generate a coherent passage that factually contradicts the correct entities in the ground-truth evidence. 

\begin{itemize}
[leftmargin=*,itemsep=0pt,topsep=0pt,parsep=0pt]

    \item \textbf{Entity substitution method}. Both NQ and HotpotQA are FQA datasets derived from Wikipedia. Following \cite{longpre2021entity}, we employ an entity substitution method in the ground-truth evidence. Using the named entity recognizer SpaCy \cite{spacy}, we categorize all ground-truth answers into five types: person, date, numeric, organization, and location, creating an entity corpus for each dataset. For every correct answer $a$, we replace all instances of $a$ in the ground-truth evidence with a different entity $a'$ randomly selected from the entity corpus. We employ both Corpus Substitution and Type Swap Substitution as described in \cite{longpre2021entity}, where $a$ and $a'$ share the same entity type or have different entity types, respectively. This process is repeated five times for each substitution type, resulting in 10 candidate passages that are highly relevant but contain incorrect answers.

    \item \textbf{Generation-based method}. For MSMARCO-QA, the answers are sentence-level, manually crafted by human annotators with provided passages \cite{nguyen2016ms}. 
    These answer sentences may be non-existent in the provided context. 
    Therefore, using a hard entity substitution approach may not be suitable. 
    Following \cite{xie2023adaptive}, we propose employing a generative approach over the correct answers for MSMARCO-QA: 
    \begin{enumerate*}[label=(\roman*)]
    
    \item For each correct answer, we pick entities not mentioned in the question and randomly choose one for both \textit{Corpus Substitution} and \textit{Type Swap Substitution}, repeating each operation five times. This results in ten incorrect answers.
    
    \item Treating each incorrect answer as a claim, we employ DeBERTa-V2  \cite{he2021deberta} to identify contradictory claims. Only the claims contradicting the answer are retained.

    \item We input the retained claims into \acp{LLM} and direct \acp{LLM} to create realistic fake evidence supporting each claim. Here, we use ChatGPT with a temperature set to 0.7. The prompt is: ``\textit{Given a claim, please write a short piece of evidence to support it. The maximum length of the generated evidence is 100 words. You can fabricate content, but it should be as realistic as possible. Claim:} \{\textit{claim}\} \textit{Evidence:}.''
    
    \item To verify that the evidence indeed supports the claim, we use an NLI model for support-checking. Specifically, the DeBERTa-V2 model determines whether the fake evidence supports the claim, and only the evidence supporting the claim is retained as the passages for subsequent experiments.
    \end{enumerate*}

\end{itemize}

\heading{Highly relevant noisy passages (HRNP)} We select original passages from the retrieved results that are highly relevant to the question but do not contain any information of the answer.  
For NQ and HotpotQA, we select 10 passages from top to bottom in the top 100 retrieval results of each question, that do not contain answer entities. 
For MSMARCO-QA, a human annotation label $is\_selected$ is included to indicate whether the passage, ranked among the top retrieved results by Bing, is used to answer the question.  
We choose passages labeled as 0 (indicating not selected as the supporting evidence) from the retrieval results.  
We regard the 10 passages organized in descending order of relevance as our HRNP.

\begin{table}[t]
  \caption{Dataset statistics.}
      \renewcommand{\arraystretch}{0.95}
  \label{tab:Data statistics}
    \begin{tabular}{l ccc}
        \toprule
        & NQ & HotpotQA & MSMARCO-QA\\
        \midrule
        \#queries  & 1863 & 4407 & 3121\\
        \#passages  & 21M & 21M & 8.8M\\
        \#ground-truth evidence & 1.0 & 2.4 & 1.1\\
        \#counterfactual passages & 3.0 & 2.4 & 2.7\\
        \#highly relevant  & 3.0 & 2.6 & 3.1\\
        \#weakly relevant & 3.0 & 2.6 & 3.1\\
       \bottomrule
    \end{tabular}
    \vspace{-2mm}
\end{table}

\begin{figure*}[t]
    \centering
    \includegraphics[width=\linewidth]{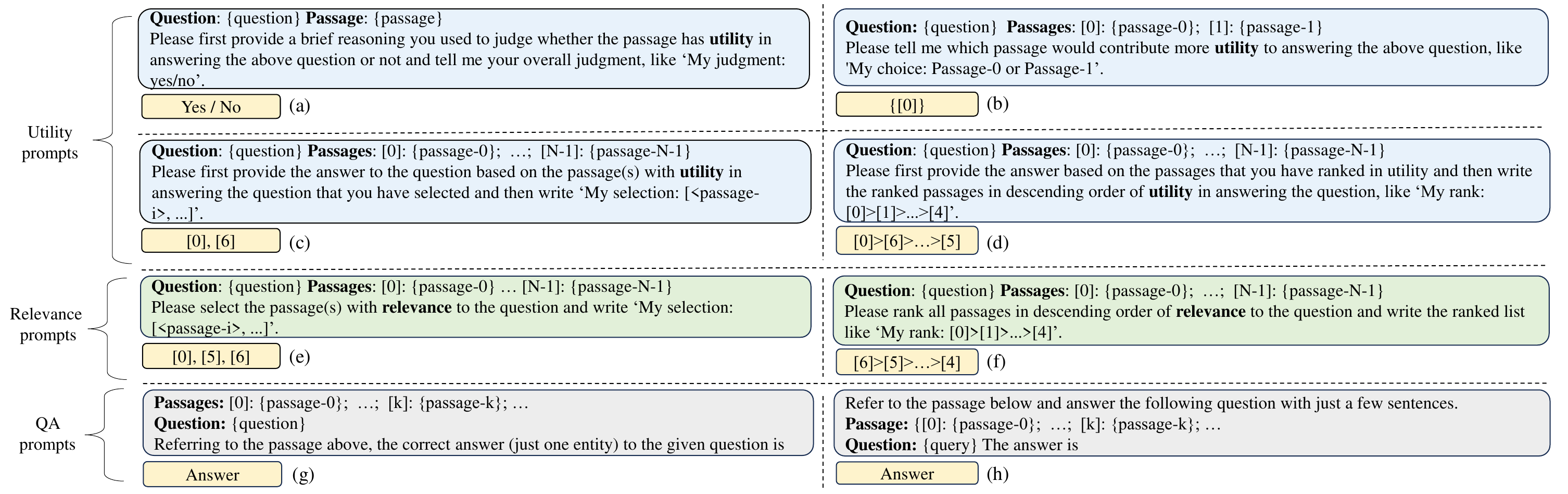}
    \vspace{-3mm}
    \caption{Prompts in blue blocks are utility prompts, where (a) is pointwise, (b) is pairwise, (c) is listwise-set, and (d) is listwise-rank.
    Prompts in green blocks are relevance prompts, where (e) is listwise-set and (f) is listwise-rank. 
    Prompts in gray blocks are QA prompts, where (g) is designed for FQA datasets and (h) is designed for NFQA dataset.}

\vspace{-1mm}
    \label{fig:prompt}
\end{figure*}

\heading{Weakly relevant noisy passages (WRNP)} We select passages from the retrieval results that are weakly relevant to the question and do not contain any information of the answer. For all datasets, we select 10 passages from bottom to top in the top 100 retrieval results of each question, that do not contain answer entities (excluding HRNP and ground-truth evidence).  
We regard the 10 passages organized in ascending order of relevance as our WRNP. 

\heading{Constructing the candidate passages} For the final set $\mathcal{D}$ with a size of 10, besides the ground-truth evidence, the passages selection follows this procedure: If CP, HRNP, and WRNP can be evenly distributed, allocate them accordingly. 
If an even distribution is not feasible, distribute as evenly as possible initially, and then randomly assign the remaining passages. 
For example, with 2 ground-truth evidence, allocate 2 to CP, HRNP, and WRNP each. The remaining 2 passages are chosen randomly: for CP, we randomly sample passages from the candidates; for HRNP and WRNP, we select passages from top to bottom.  
The statistics are presented in Table \ref{tab:Data statistics}.

\vspace{-2mm}
\subsection{Instructing \acp{LLM} with prompts}
\label{prompts}

We introduce the \acp{LLM} used for evaluation and the prompts used for guiding \acp{LLM}. 
We design three types of prompts, i.e., utility prompts and relevance prompts for evidence selection, and QA prompts for answer generation; see Fig.~\ref{fig:prompt}. 

\heading{\acp{LLM} for evaluation}
We have selected several representative closed-source  and open-source \acp{LLM} for our analysis:
\begin{enumerate*}[label=(\roman*)]
    \item{Closed-source \acp{LLM}} For the closed-source \acp{LLM}, we conduct our experiments using OpenAI's API \cite{chatgpt}, specifically gpt-3.5-turbo-1106 (abbreviated as ChatGPT). 

    \item{Open-source \acp{LLM}} As for the open-source \acp{LLM}, our experiments involve four models: Llama2-7B-chat (abbreviated as Llama2-7B) \cite{touvron2023llama}, Llama2-13B-chat (abbreviated as Llama2-13B) \cite{touvron2023llama}, Vicuna-7B \cite{chiang2023vicuna}, and Vicuna-13B \cite{chiang2023vicuna}.
\end{enumerate*}

\heading{Utility prompts design}
The goal of utility prompts is to guide \acp{LLM} to select evidence with utility in answering the question. 
According to different input forms, we consider three ways of presenting candidate passages as input to \acp{LLM}:
\begin{enumerate*}[label=(\roman*)]
    \item \textbf{Pointwise}: Each candidate passage $d \in \mathcal{D}$ is individually concatenated with the question $q$, and $N$ such inputs are separately presented to \acp{LLM}. If the \acp{LLM} output $yes$ (Fig.~\ref{fig:prompt}(a)), this indicates the passage's utility in addressing the question. 
    
    \item \textbf{Pairwise}: Two candidate passages, $d_i \in \mathcal{D}$ and $d_j \in \mathcal{D}$, are concatenated with $q$. If \acp{LLM} output $d_i$ (Fig.~\ref{fig:prompt}(b)), this suggests that \acp{LLM} find that $d_i$ contributes more utility than $d_j$ in answering the question. This process iterates $N(N-1)/2$ times, yielding the overall ranking of supporting evidence. 
    \item \textbf{Listwise}: All candidate passages $\{d_1, d_2,\ldots, d_N\}$ in $\mathcal{D}$ are concatenated with $q$. Two output formats are designed:
    \begin{enumerate*}[label=(\roman*)]

        \item  the set of evidence with utility (\textbf{Listwise-set}, Fig.~\ref{fig:prompt}(c)), and
    
        \item the evidence list ranked by utility (\textbf{Listwise-rank}, Fig.~\ref{fig:prompt}(d)).
    \end{enumerate*}
\end{enumerate*}




    

\begin{figure*}[t]    
\centering
    \includegraphics[width=\linewidth]{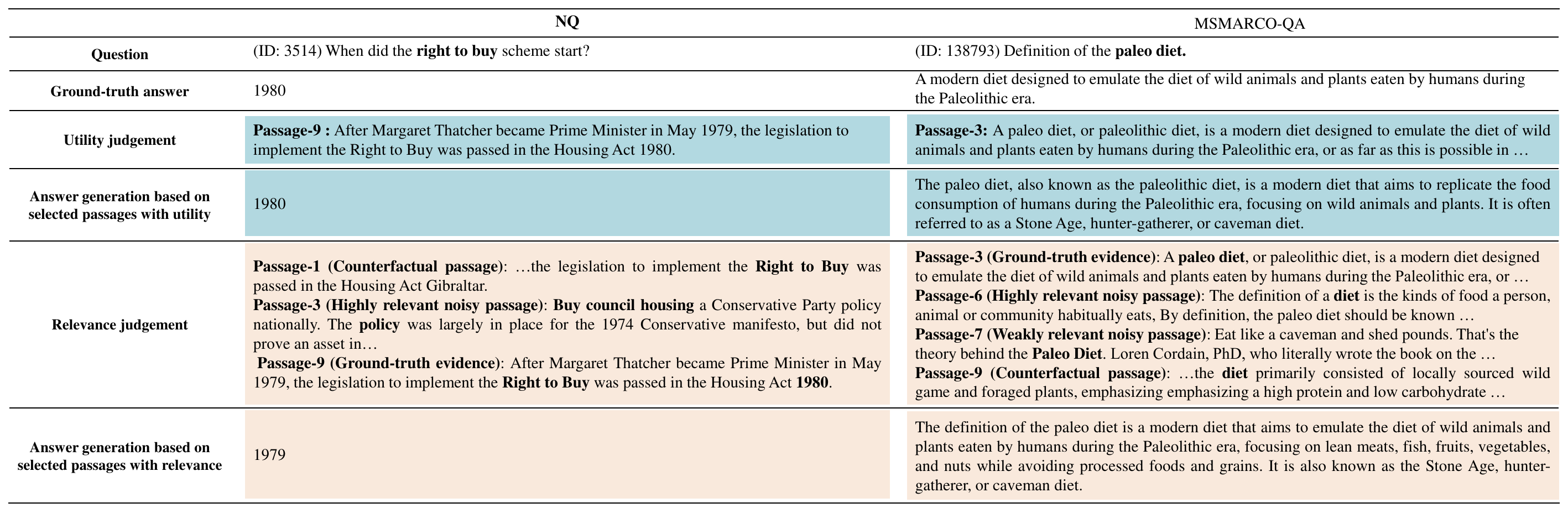}
    \caption{Given 10 candidate passages, ChatGPT is employed to select evidence with utility and relevance respectively using listwise-set approaches. The selected evidence are then used by ChatGPT to answer questions. In the examples from NQ and MSMARCO-QA, ``Passage-9'' and ``Passage-3'' respectively denote the ground-truth supporting evidence. The full set of 10 candidate passages for each question can be accessed at \url{https://github.com/ict-bigdatalab/utility_judgments}.}
    \label{fig:relevance_utility_exp}
    \vspace{-1mm}
\end{figure*}

\heading{Relevance prompts design}
The goal of relevance prompts is to guide LLMs to select evidence that are relevant to the question. 
Following \cite{sun2023chatgpt}, we only consider listwise input. Similar to utility prompts, we also account for two distinct outputs: listwise-set and listwise-rank, as elaborated in Fig.~\ref{fig:prompt}(e) and~\ref{fig:prompt}(f). 

\heading{QA prompts design}
For GTU, since the presence of ground-truth evidence is uncertain, directly assessing the chosen evidence by \acp{LLM} becomes challenging. 
Consequently, we design QA prompts to evaluate their QA abilities, which  guide \acp{LLM} to obediently answer the questions based on evidence with utility or relevance. 
As depicted in Fig.~\ref{fig:prompt}(g) and~\ref{fig:prompt}(h), QA prompts are crafted for the FQA and NFQA datasets, respectively.

\vspace{-2mm}
\subsection{Evaluation metrics} \label{eva_metrics}

\textbf{GTI evaluation.} 
We assess the performance of \acp{LLM} in selecting evidence for GTI, categorizing the output format into two types: 
\begin{enumerate*}[label=(\roman*)]

    \item \textbf{The evidence set}. For pointwise and listwise-set approaches, the final output of \acp{LLM} is a set of evidence based on utility or relevance. We employ \textbf{Precision} (\textbf{P}), \textbf{Recall} (\textbf{R}) and \textbf{F1}.

    \item \textbf{The evidence list}. For pairwise and listwise-rank approaches, the output of \acp{LLM} is a ranked list based on utility or relevance. 
    We use widely-used evaluation metrics in IR, i.e., mean reciprocal rank (\textbf{MRR}) \cite{craswell2009mean} and normalized discounted cumulative gain (\textbf{NDCG}) \cite{jarvelin2017ir}.

\end{enumerate*}

\heading{GTU evaluation} 
We consider the final answering performance based on selected evidence by \acp{LLM} for GTU. 
Following \cite{izacard2020leveraging, ren2023investigating}, we use the exact match (\textbf{EM}) score and \textbf{F1} score to evaluate the answer performance of \acp{LLM} on the FQA datasets. 
Following \cite{nguyen2016ms}, we use \textbf{ROUGE-L} \cite{lin2004rouge} and \textbf{BLEU} \cite{papineni2002bleu} to evaluate the answer performance of \acp{LLM} on the NFQA dataset. 

\vspace{-2mm}
\section{Relevance vs. Utility}\label{exp-1}

We use ChatGPT as an example to investigate whether \acp{LLM} can distinguish between utility and relevance (\textit{RQ1}).

\heading{Experimental setup} We evaluate ChatGPT \cite{chatgpt} using the NQ, HotpotQA, and MSMARCO-QA datasets incorporated into the GTI benchmark. 
We employ the listwise forms for both relevance and utility judgments, including listwise-set in Fig.~\ref{fig:prompt}(c) and~\ref{fig:prompt}(e), and listwise-rank in Fig.~\ref{fig:prompt}(d) and~\ref{fig:prompt}(f).  
For each question, with 10 candidate passages, we shuffle them as the input and the input passages for relevance judgments and utility judgments remain the same. 

\begin{table}[t]
\small
  \centering
  \caption{The performance (\%) of utility judgments and relevance judgments using ChatGPT under the listwise approaches on the GTI benchmark.}
  \renewcommand{\arraystretch}{0.95}
   \setlength\tabcolsep{3.2pt}
    \begin{tabular}{l l ccc ccc}
    \toprule
  \multicolumn{1}{l}{\multirow{3}[6]{*}{Dataset}} & \multicolumn{1}{c}{\multirow{3}[6]{*}{Judgment}} & \multicolumn{3}{c}{Listwise-set} & \multicolumn{3}{c}{Listwise-rank} \\
\cmidrule(r){3-5}   \cmidrule(r){6-8}       &       & \multicolumn{1}{c}{P} & \multicolumn{1}{c}{R} & \multicolumn{1}{c}{F1} & \multicolumn{2}{c}{NDCG} & \multicolumn{1}{c}{MRR} \\
\cmidrule(r){6-7}   \cmidrule(r){8-8}       &       &       &       &       & @1     & @5     & @5 \\
      \midrule
   \multirow{2}{*}{NQ} & Relevance & 36.65 & 73.75   & 48.97  & 39.08   & 67.62   & 59.42      \\
   & Utility  & 57.19 & 73.00 & 64.14  & 57.80    &  77.43  & 71.87  \\ 
    \midrule
    \multirow{2}{*}{HotpotQA} & Relevance & 70.02  &  47.45 & 56.56 & 74.54 &  79.67   &  85.75   \\
   & Utility & 76.83  & 46.54 & 57.97 & 78.28  & 80.29   & 87.52\\
    \midrule
    \multirow{2}{*}{MSMARCO-QA} & Relevance & 32.80 & 85.04 & 46.87  & 40.82  &  66.07   & 60.17 \\
    & Utility & 36.77  & 63.50 & 46.57  & 40.07   & 64.90   & 60.35 \\
    \bottomrule
    \end{tabular}%
  \label{tab:relevance_utility_tab}%
\end{table}%

\heading{\acp{LLM} can distinguish between utility and relevance} 
Table \ref{tab:relevance_utility_tab} presents a comparative analysis of ChatGPT's performance in relevance and utility judgments across three datasets. 
Utility-based prompts prove more effective in assisting \acp{LLM} in identifying ground-truth evidence compared to relevance-based prompts, particularly in NQ dataset. 
E.g., in NQ, the F1 and NDCG@1 scores for utility judgments exhibit a notable increase of 30.98\% and 47.90\%, respectively, compared to relevance judgments.

\heading{\acp{LLM} exhibit distinct performance on multi-hop and single-passage QA datasets} 
\begin{enumerate*}[label=(\roman*)]
    \item In the listwise-rank approach, the performance of utility judgments and relevance judgments is superior on the multi-hop dataset, i.e.,  HotpotQA, compared to the single-passage QA dataset, e.g., NQ. This could be attributed to the presence of multiple pieces of ground-truth evidence for each question in HotpotQA, leading to a higher probability of the ground-truth evidence appearing at the top of the ranked list. 
    \item In the listwise-set approach, the performance of relevance judgments on HotpotQA surpasses that on NQ in terms of F1, likely due to the increased probability of the set containing ground-truth in HotpotQA. 
    However, concerning utility judgments, the F1 score for HotpotQA is lower than for NQ. This discrepancy may arise from the \acp{LLM}'s capability to address multi-hop questions, where they may not recall all necessary evidence required at each step, consequently impacting their judgment, particularly when selecting precise sets.
\end{enumerate*}

\begin{table*}[t]
  \centering
  \small
  \caption{The performance (\%) of different \acp{LLM} in utility judgments under different datasets and input forms. 
  Bold indicates the best performance among the same type of input. }
    \renewcommand{\arraystretch}{0.95}
   \setlength\tabcolsep{3.3pt}
    \begin{tabular}{l ccc ccc ccc }
    \toprule
     \multicolumn{10}{c}{\textit{Pointwise} / \textit{Listwise-set}}  \\
     \midrule
       \multirow{2}{*}{Model}  &  \multicolumn{3}{c}{NQ} &  \multicolumn{3}{c}{HotpotQA} & \multicolumn{3}{c}{MSMARCO-QA} \\    
     \cmidrule(r){2-4} \cmidrule(r){5-7} \cmidrule(r){8-10}  &   Precision   & Recall  & F1  & Precision   & Recall  & F1  & Precision   & Recall  & F1  \\
  \midrule
    ChatGPT
    & \textbf{22.40}  / \textbf{57.19}  &  \phantom{1}92.97  / \textbf{73.00}    &    \textbf{41.00}  / \textbf{64.14} &   \textbf{46.04}   / \textbf{76.83}     &   46.07  / 46.54     &    46.05   / \textbf{57.97}   &     \textbf{25.26}  / \textbf{36.77}     &  73.82  /  \textbf{63.50}        &     \textbf{37.64}  / \textbf{46.57}  \\

    LlaMa2-7B 
     &  18.25 / 15.46 &  \phantom{1}81.27  / 39.56  & 29.80 / 22.23      &  33.97 / 49.13   & 48.59 / 35.83   & 39.98 / 41.44 &  14.41 / 21.84  &     89.78 / 41.40  & 24.83 / 28.59   \\

    LlaMa2-13B  
     &  17.69 / 27.56  & \phantom{1}36.71 / 38.43  &   23.88 / 32.10    & 26.71 / 61.49    &  20.68 / 36.80   & 23.31  / 46.04  &  16.23 / 23.15  &   48.40 / 27.45     & 24.31 / 25.12  \\
   Vicuna-7B 
    &  13.39 / 18.67  & \textbf{100.00}  / 45.14    &  23.62  / 26.42       & 37.27  / 44.46   & \textbf{89.76} / 45.77   & \textbf{52.67} / 45.10   &  13.23 / 18.41  &  \textbf{96.87} / 35.27     &  23.29 / 24.19 \\

    Vicuna-13B
     &  15.02 / 35.24  & \phantom{1}86.96 / 66.02    &   25.62 / 45.95    &  36.75 / 60.23     & 76.77  / \textbf{46.72}  &  49.70 / 52.62   &  14.39 / 24.62  &   85.91 / 53.84       &   24.66 / 33.79    \\
     \midrule
    \multicolumn{10}{c}{\textit{Pairwise} / \textit{Listwise-rank} }  \\
     \midrule
      \multirow{2}{*}{Model}  &  \multicolumn{3}{c}{NQ} &  \multicolumn{3}{c}{HotpotQA} & \multicolumn{3}{c}{MSMARCO-QA}  \\ 
    \cmidrule(r){2-4} \cmidrule(r){5-7} \cmidrule(r){8-10}    & NDCG@1  & NDCG@5  & MRR@5  & NDCG@1  & NDCG@5 & MRR@5  & NDCG@1  & NDCG@5 & MRR@5 \\
  \midrule
    ChatGPT
   &    \textbf{57.11} / \textbf{57.80}   &    \textbf{78.23} / \textbf{77.43}    &      \textbf{72.18} / \textbf{71.87} &   \textbf{74.67} / \textbf{78.28}    &  \textbf{75.11} / \textbf{80.29}  &    \textbf{84.94} / \textbf{87.52}  &     \textbf{29.89} / \textbf{40.07}   &   \textbf{57.56} / \textbf{64.90}     &    \textbf{50.63} / \textbf{60.35}  \\

    LlaMa2-7B  
     &   13.53 / \phantom{1}7.66 & 34.99 / 23.12 & 27.69 / 17.86    & 29.18 / 24.03 & 42.33 / 32.85 & 49.21 / 38.99 & 11.49 /  \phantom{1}4.69     & 32.85  / 11.48    & 26.76 / \phantom{1}9.45     \\

    LlaMa2-13B
     & 16.96 / \phantom{1}6.93  &  39.03 / 12.52   & 31.66  / 10.95     &  31.02 / 20.10   & 42.63  / 20.22  & 50.34 / 24.78    &  14.64  / \phantom{1}6.89   & 35.20  / 14.07     &  29.49  / 12.21   \\
  
    Vicuna-7B
     & 10.03 / 10.22    &  30.18  / 29.47   &  23.37 / 23.12   &  26.11 / 27.66    & 39.49 / 37.75   &  46.24 / 45.79   & 11.22 / 12.14    & 30.59 / 29.93     &  24.97 / 25.04   \\

    Vicuna-13B 
     & 12.37 / 34.93  & 33.00  / 57.79     & 25.93  / 51.59   &  29.45  / 67.62   &  41.30 / 66.36    & 48.71  / 78.85    & 12.27 / 22.91    &  31.46 / 44.45    &  25.91  / 38.87  \\
    
    \bottomrule
    \end{tabular}%
  \label{tab:results-2}%
\end{table*}%

\heading{\acp{LLM} are highly receptive to generated counterfactual passages} 
\begin{enumerate*}[label=(\roman*)]
    \item In the listwise-set and listwise-rank scenarios, ChatGPT's performance on MSMARCO-QA is significantly worse than on NQ in terms of utility judgments. 
    This disparity may stem from the construction of counterfactual passages \cite{xie2023adaptive}.
    In NQ, counterfactual passages are built through entity substitution, potentially leading to passage incoherence. 
    \acp{LLM} might be sensitive to incoherent passages, resulting in their rejection during utility judgments. 
    Conversely, the construction of counterfactual passages in MSMARCO-QA involves \acp{LLM} generating coherent passages, which may confuse the utility judgments of \acp{LLM}. 
    \item However, for relevance judgments the performance gap between MSMARCO-QA and NQ is small. 
    Both counterfactual passages and entity substitution are highly relevant to the question, so the performance of selected passages remains low regardless of different construction approaches. 
\end{enumerate*}

\heading{Case study} Fig.~\ref{fig:relevance_utility_exp} illustrates two examples based on utility and relevance judgments. 
The evidence selected by \acp{LLM} based on utility is more precise than that selected based on relevance. 
When generating answers using utility judgments results and relevance judgments results, respectively, it becomes evident that the noise or misinformation in the relevance results significantly impacts the \acp{LLM}'s answer generation. E.g., the ``Passage-9'' obtained from relevance judgments contains misinformation about the focus of paleo diet, which misguides the answer generator's understanding of the paleo diet.
This underscores the importance of enhancing the quality of supporting evidence for answer generation.


\vspace{-2mm}
\section{Utility judgments depend on instruction design} \label{exp-2}
Our analysis suggests that utility judgments may offer more effective guidance to \acp{LLM} in identifying ground-truth evidence for answering questions. 
In light of this, we extend our analysis to explore how different factors affect utility judgments (\textit{RQ2}). 
Specifically, we examine key factors in instruction design, including the input form of passages (i.e., pointwise, pairwise, and listwise), the sequence of input between the question and passages, and additional requirements  (i.e., chain-of-thought, reasoning, and providing answers). 
We evaluate ChatGPT, Llama2-7B, Llama2-13B, Vicuna-7B, and Vicuna-13B, on the NQ, HotpotQA, and MSMARCO-QA datasets incorporated into the GTI benchmark. 

\noindent%
\textbf{Different input forms have varying impacts on utility judgments.} 
Table \ref{tab:results-2} shows the performance of pointwise, pairwise, and listwise inputs. 
We directly use the prompt in Fig.~\ref{fig:prompt} and the position of ground-truth evidence is random in the input passage list. 
\begin{figure}[t]    
\centering
    \includegraphics[width=\linewidth]{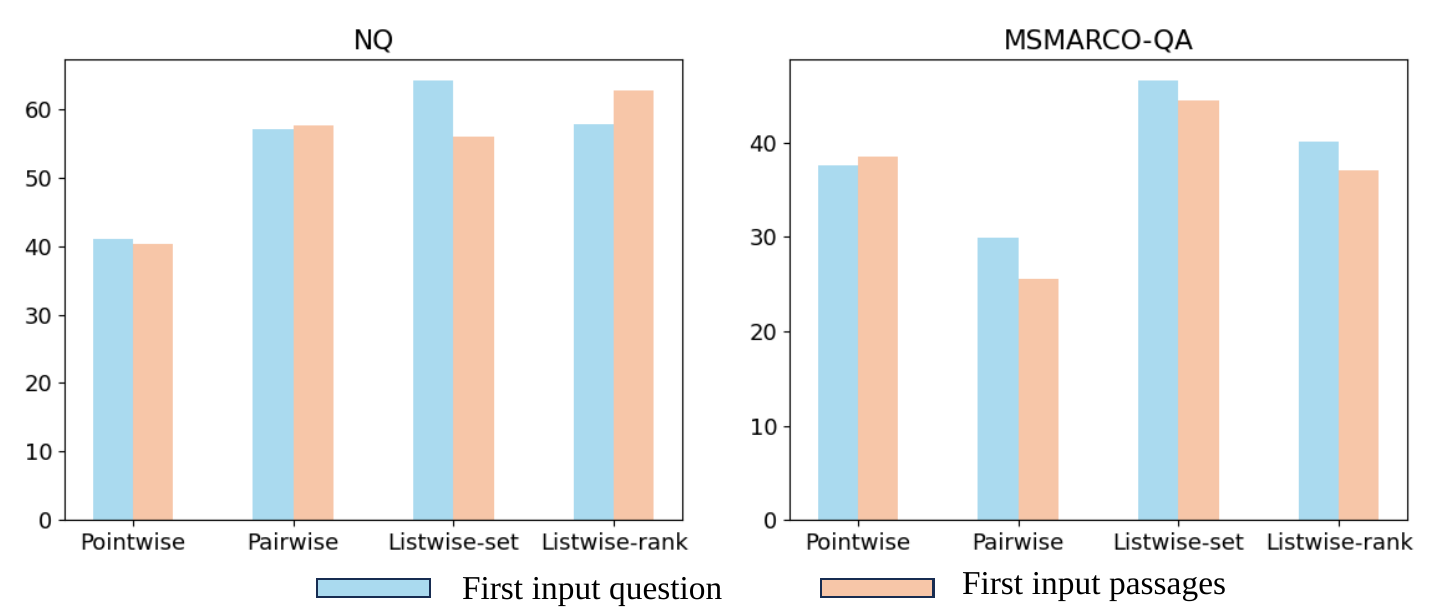}
    \caption{Performance of ChatGPT with different input forms on the NQ and MSMARCO-QA datasets in different sequences of input between the question and passages. We use ``F1'' score on pointwise and listwise-set forms and ``NDCG@1'' score on pairwise and listwise-rank forms. ``NDCG@5'' has same trend with `NDCG@1''.}
    \label{fig:question_passages_order}
    \vspace{-1mm}
\end{figure}
\begin{figure*}[t]
    \includegraphics[width=\linewidth]{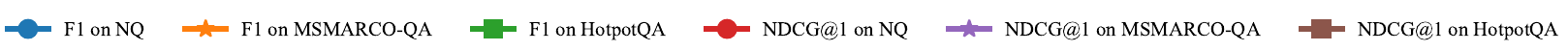}
    \\
    \vspace*{-1mm}
    \subfigure[ChatGPT (Listwise-set)]{
        \includegraphics[width=3.3cm]{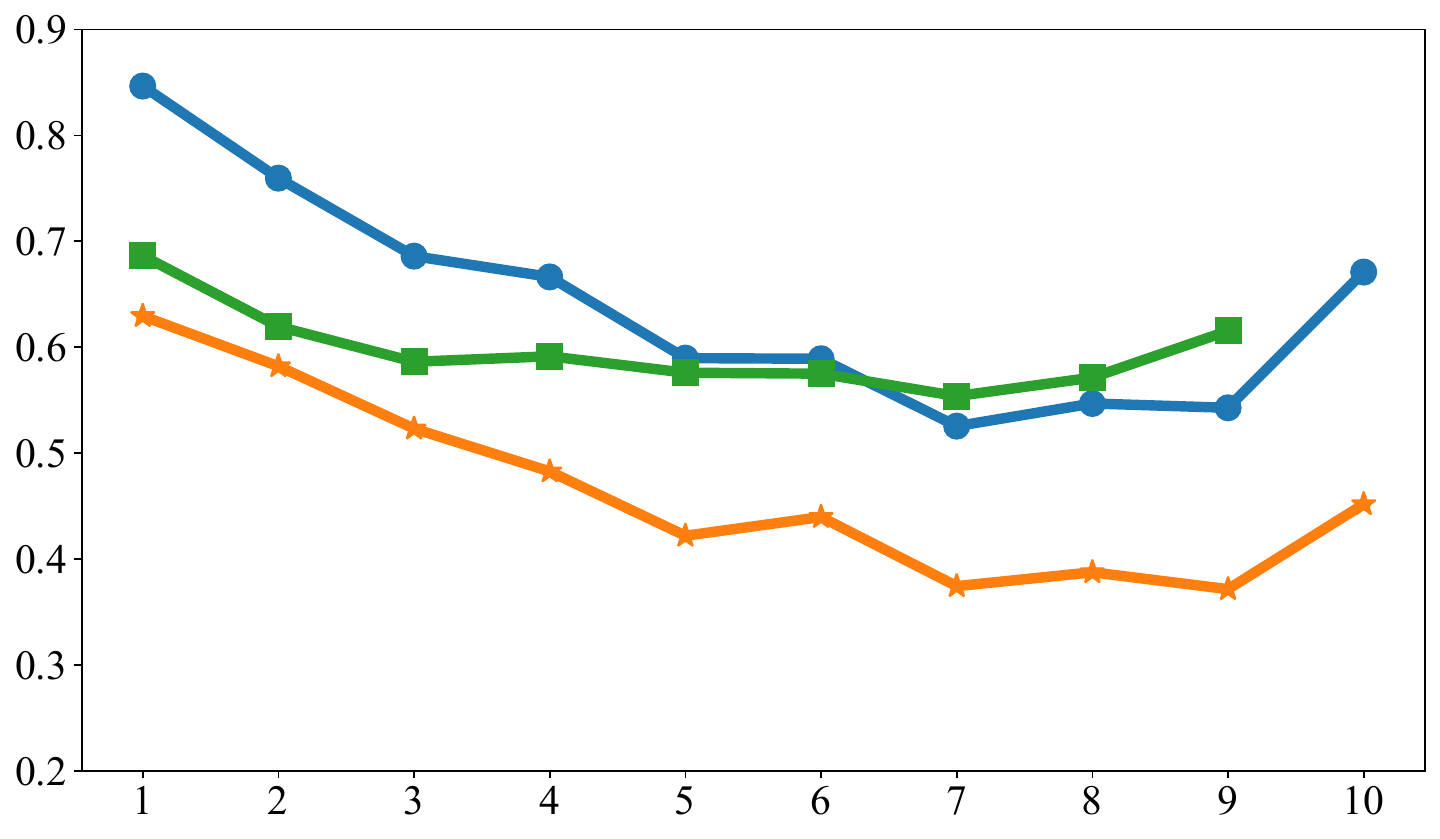}
    }
    \subfigure[LlaMa2-13B (Listwise-set)]{
        \includegraphics[width=3.3cm]{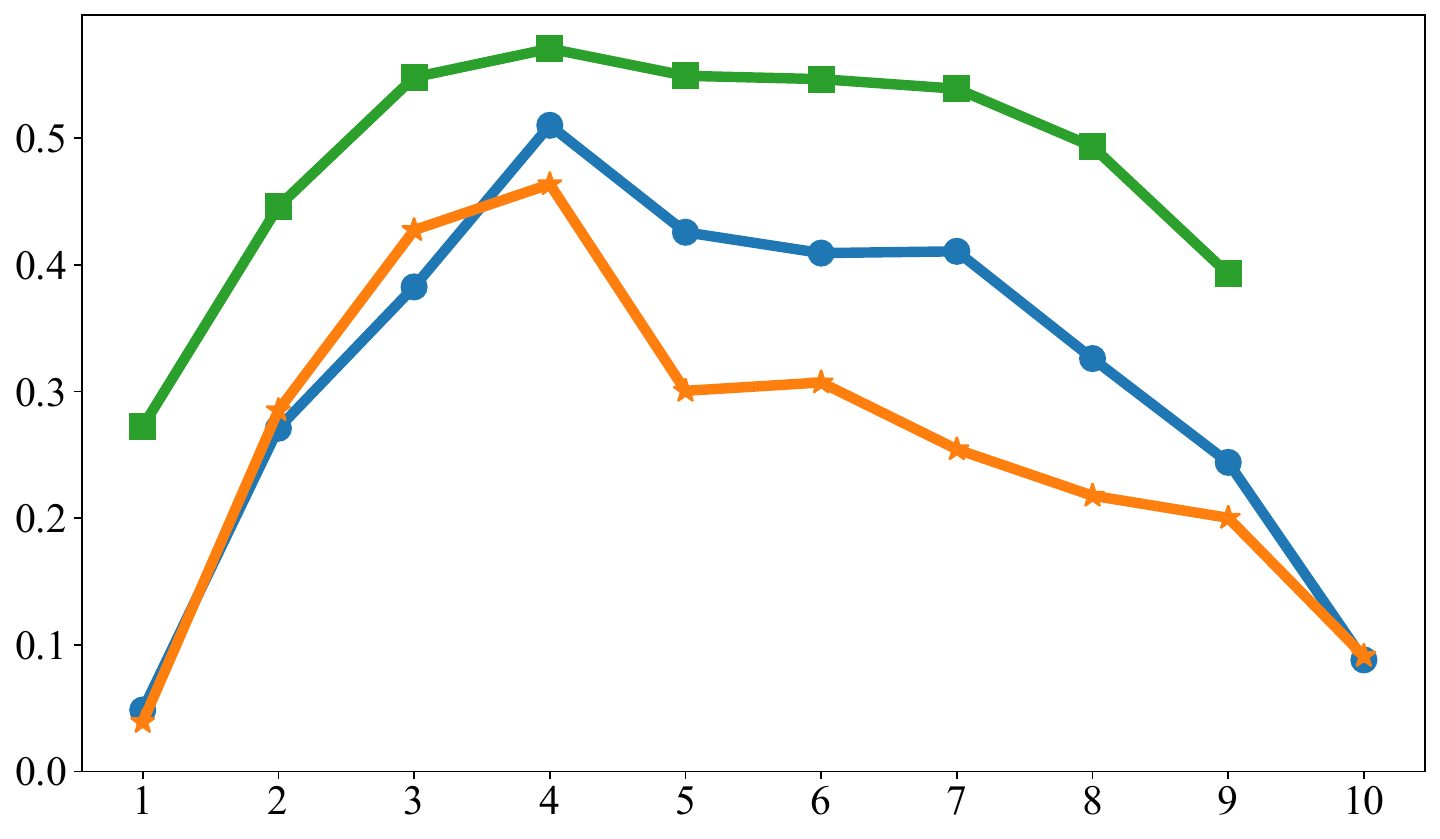}
    }
    \subfigure[Vicuna-13B (Listwise-set)]{
        \includegraphics[width=3.3cm]{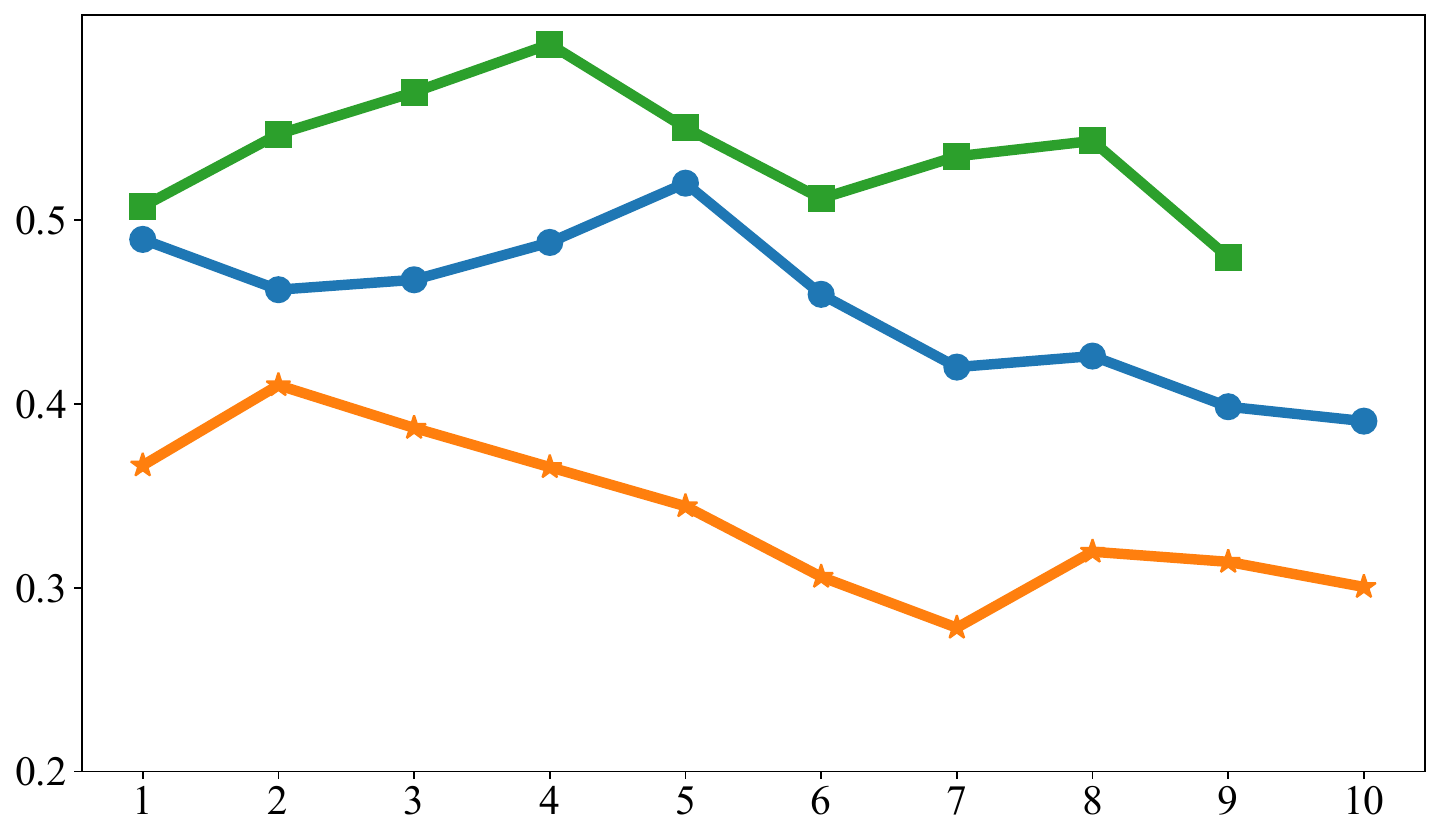}
    }
    \subfigure[LlaMa-7B (Listwise-set)]{
        \includegraphics[width=3.3cm]{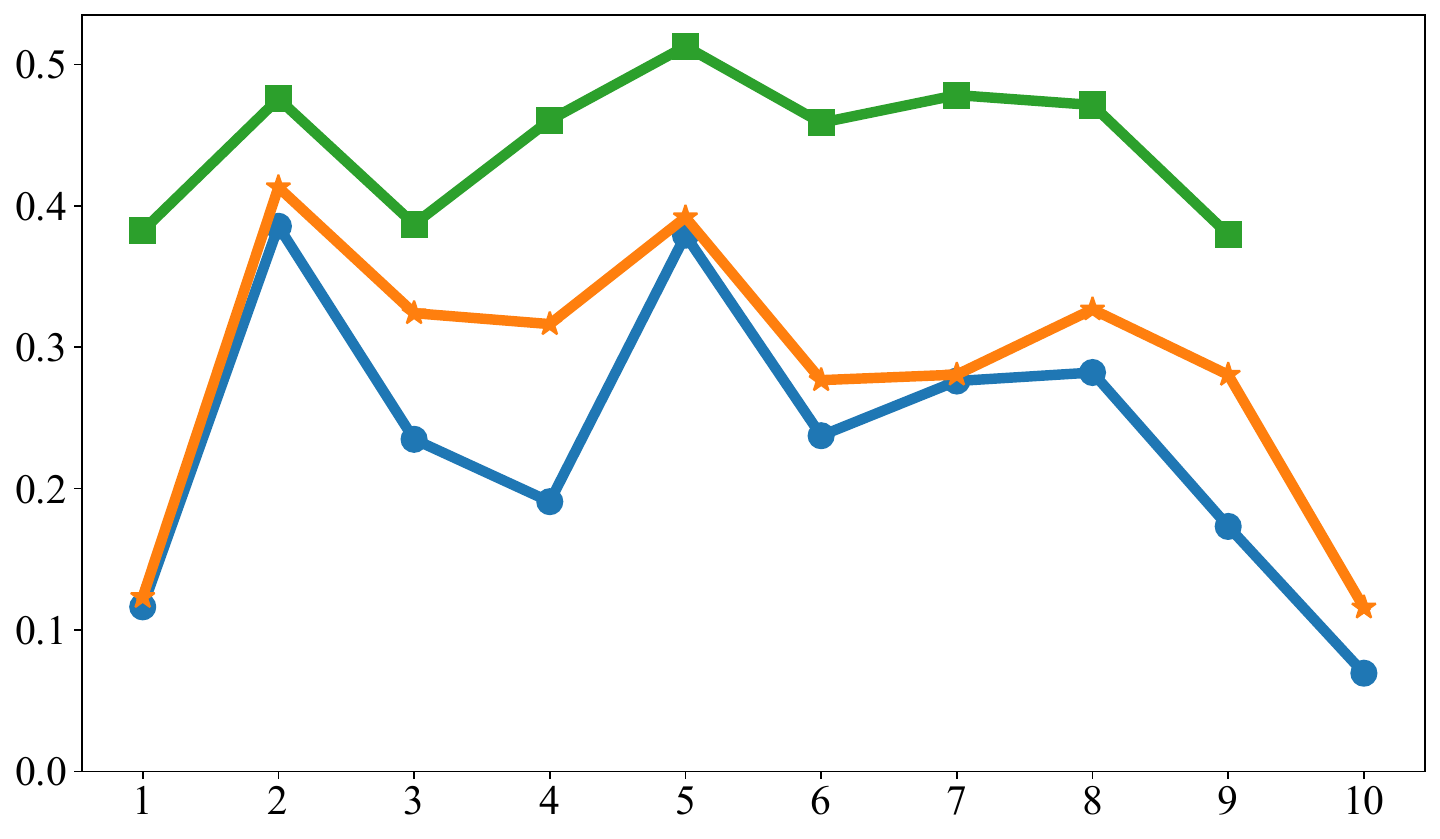}
    }
    \subfigure[Vicuna-7B (Listwise-set)]{
        \includegraphics[width=3.3cm]{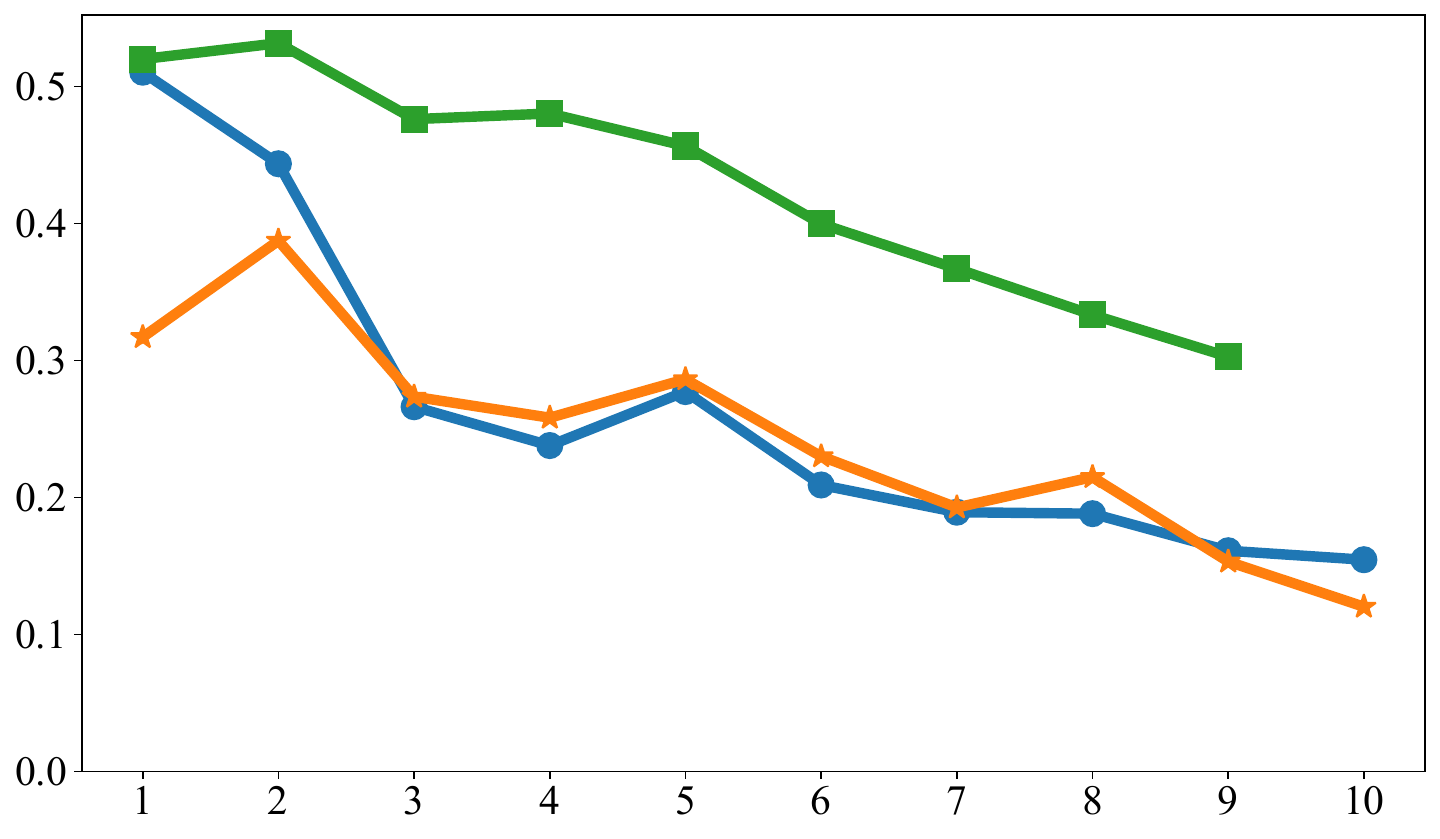}
    }
    \subfigure[ChatGPT (Listwise-rank)]{
        \includegraphics[width=3.3cm]{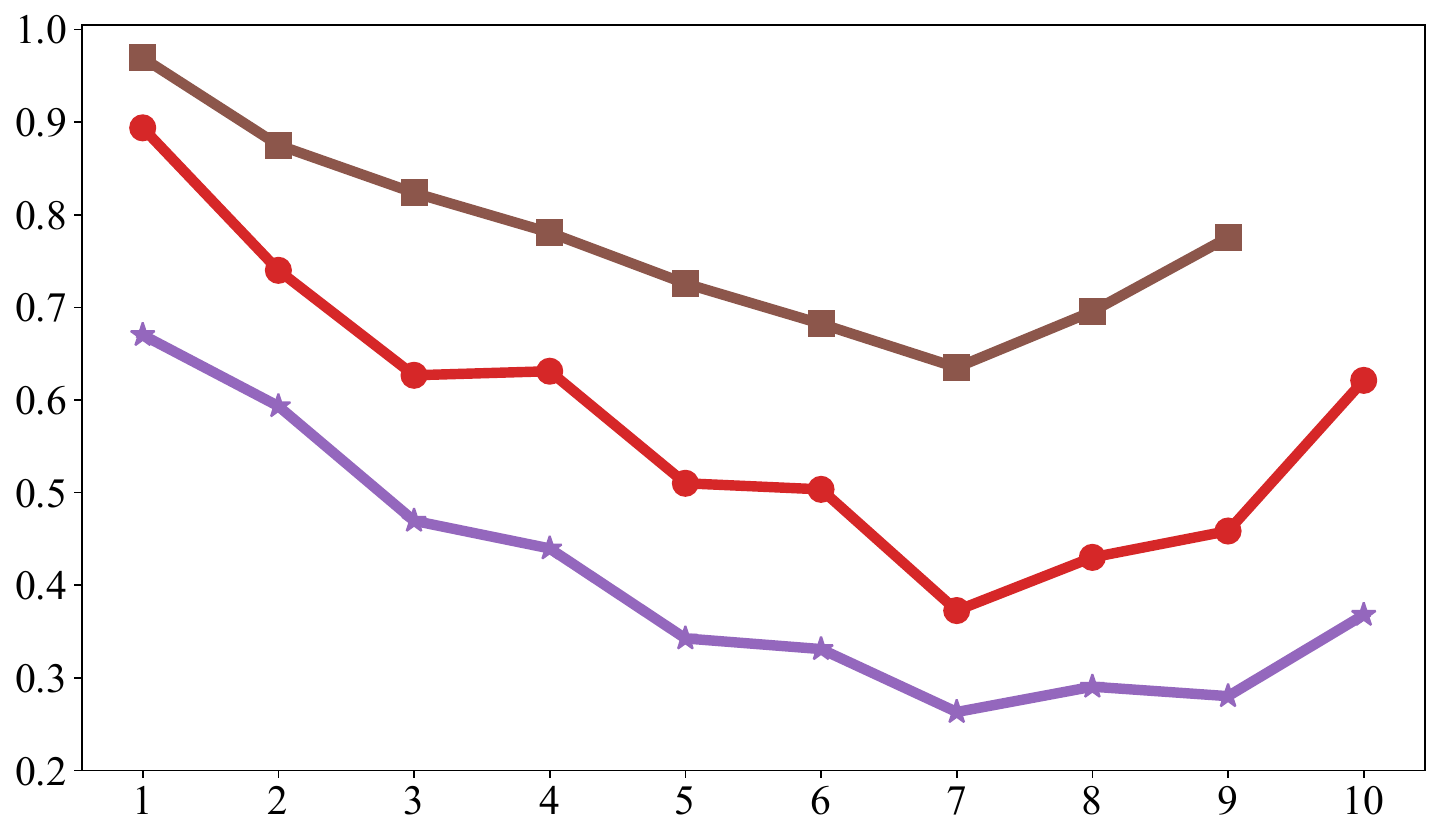}
    }
    \subfigure[LlaMa2-13B (Listwise-rank)]{
        \includegraphics[width=3.3cm]{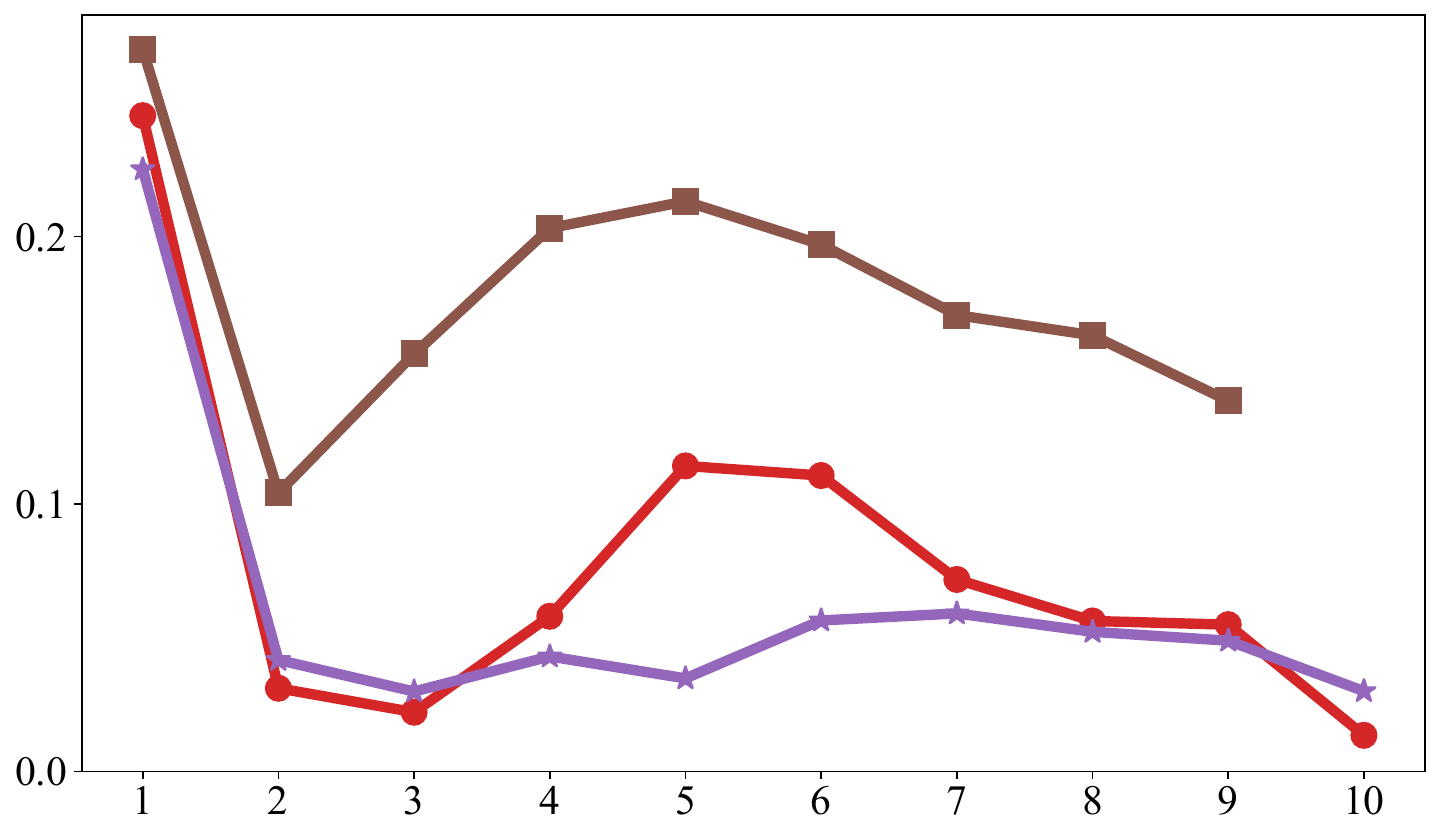}
    }
    \subfigure[Vicuna-13B (Listwise-rank)]{
        \includegraphics[width=3.3cm]{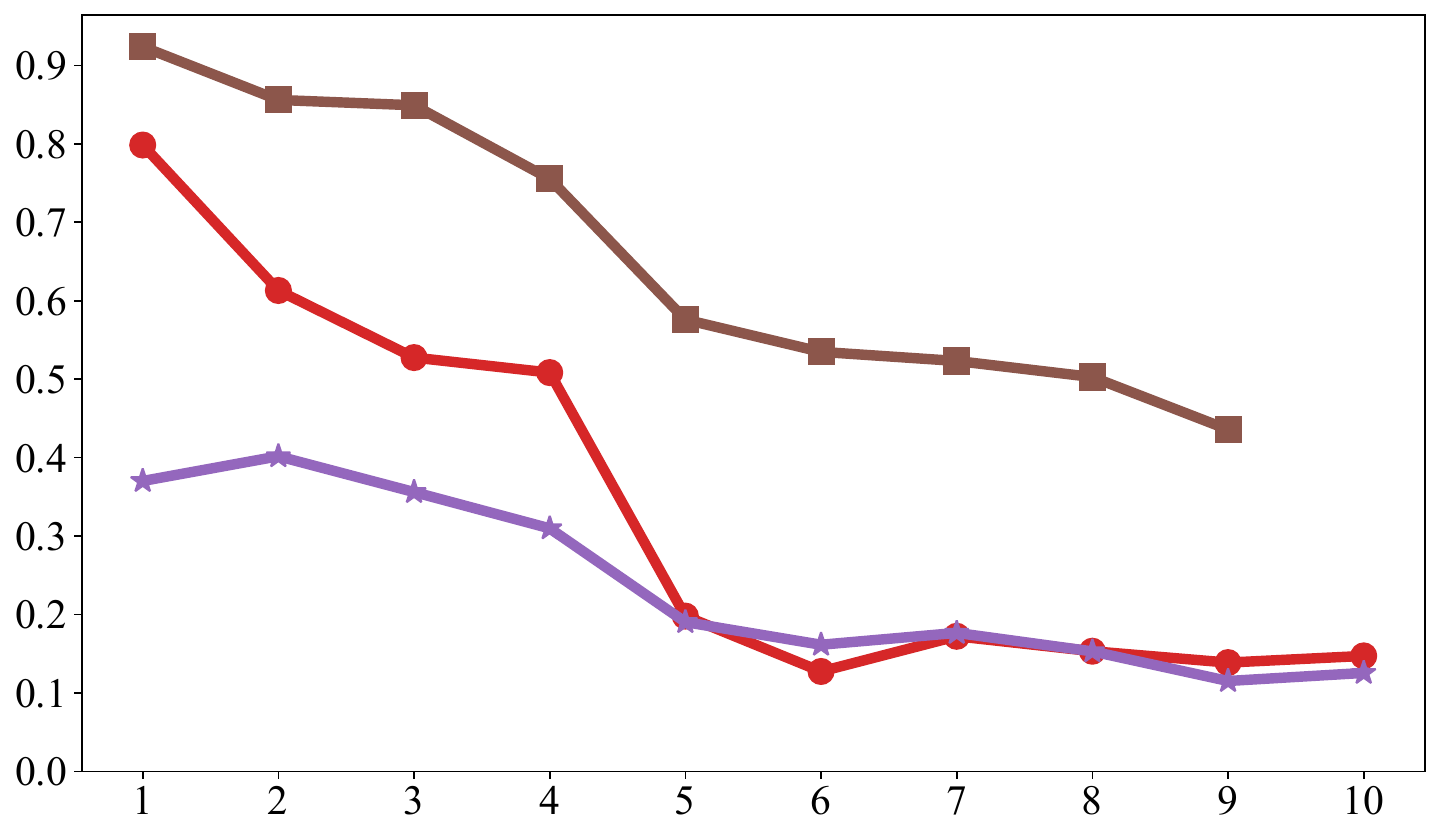}
    }
    \subfigure[LlaMa-7B (Listwise-rank)]{
        \includegraphics[width=3.3cm]{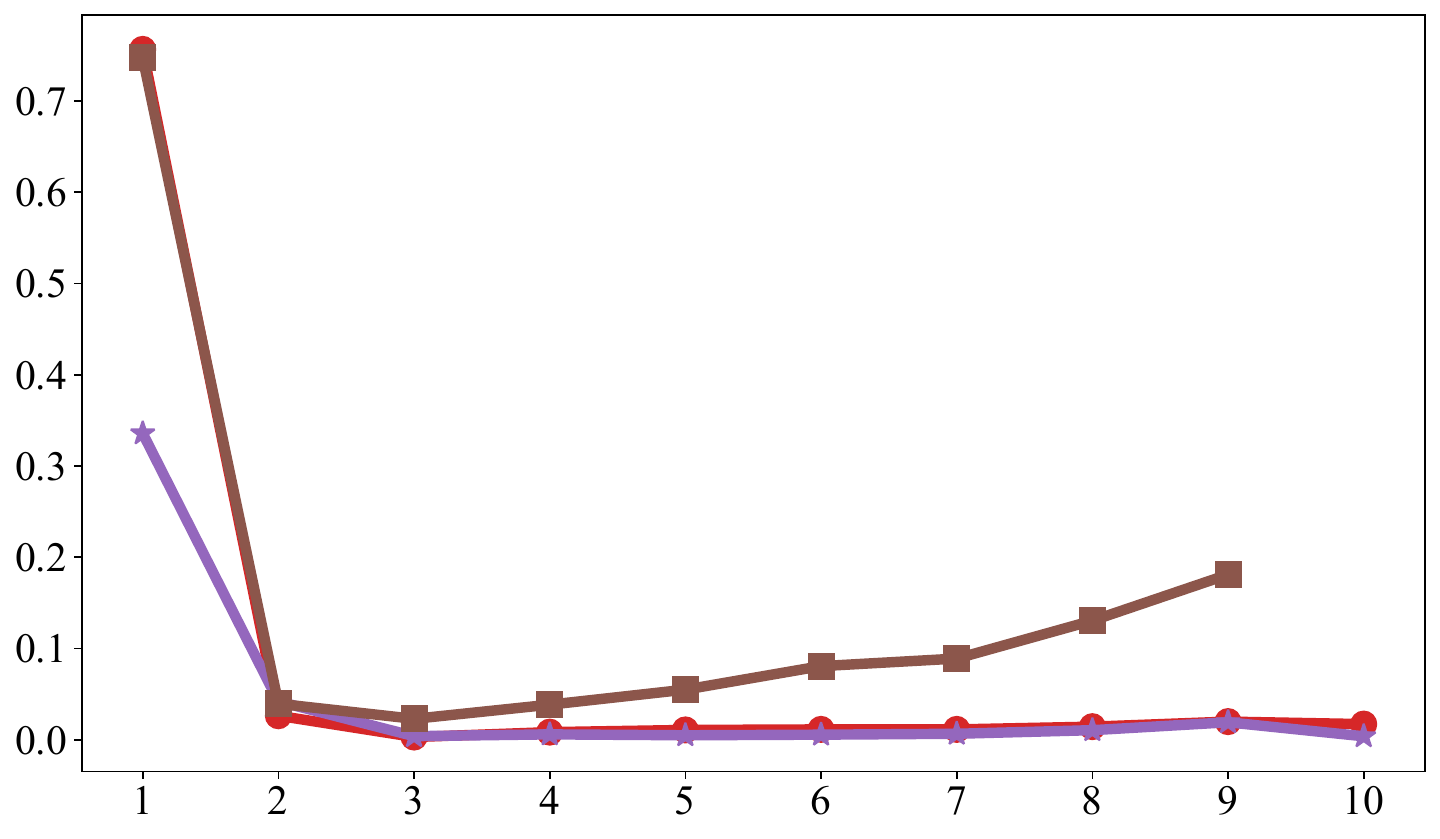}
    }
    \subfigure[Vicuna-7B (Listwise-rank)]{
        \includegraphics[width=3.3cm]{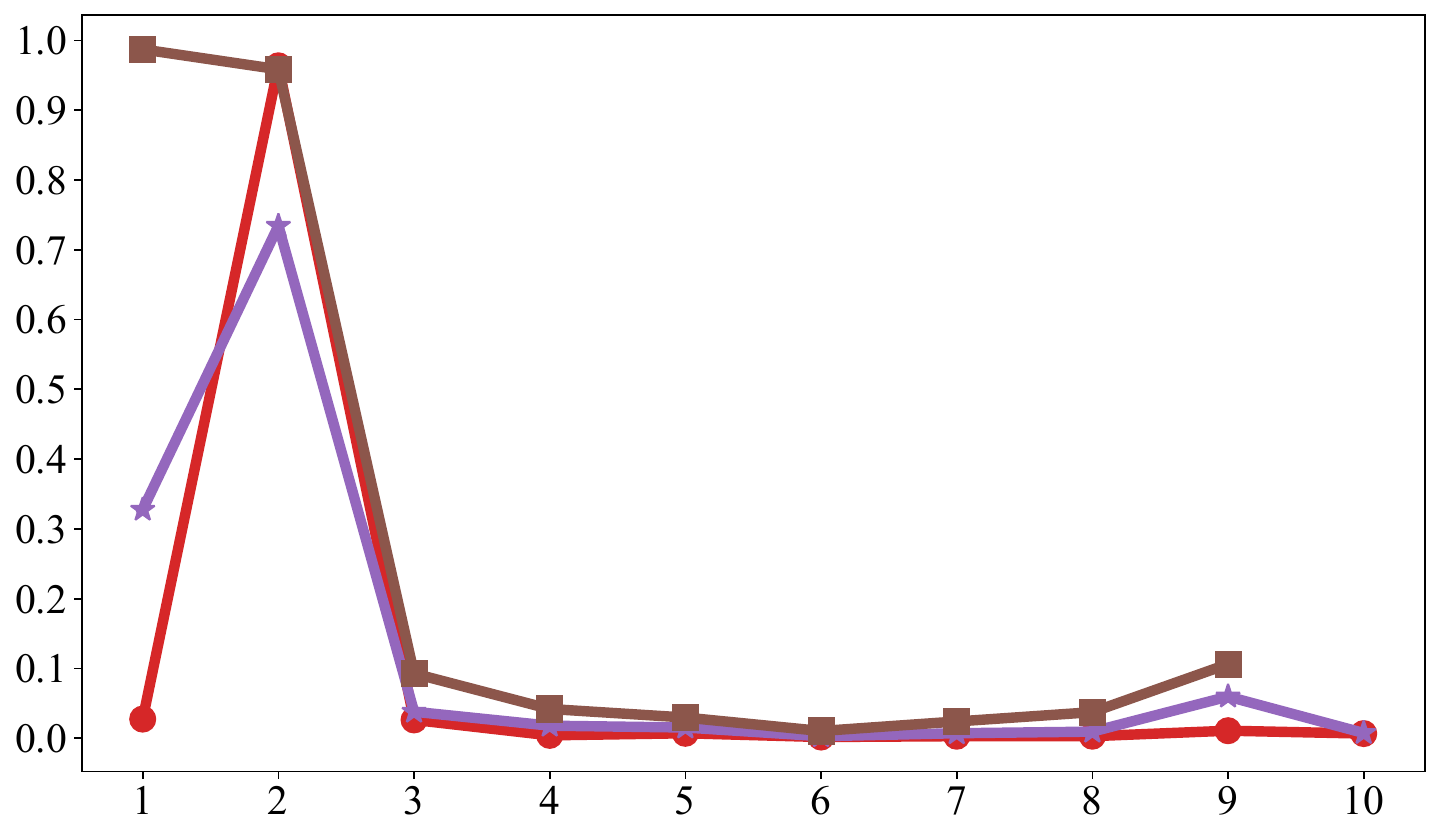}
    }
\caption{The performance of five different \acp{LLM} in utility judgments based on the positions of ground-truth evidence in the input list across listwise-set and listwise-rank forms on different datasets.}
\label{fig:positions}       
\vspace{-2mm}
\end{figure*}  

    \textbf{Pointwise}. \acp{LLM} show high recall but low precision, leading to low F1 across all datasets. Analyzing \acp{LLM} outputs reveals a tendency, except for Llama2-13B, to frequently output ``yes.''

    \textbf{Pairwise}. 
        For the same \acp{LLM}, the performance of utility judgments is better on the NQ and HotpotQA datasets. There is room for improvement in the ability to perform utility judgments on the constructed MSMARCO-QA dataset. The reasons could be twofold. Firstly, the MSMARCO-QA dataset includes numerous non-factual questions and might impact utility judgments capabilities of the \acp{LLM} compared to NQ. Secondly, the input passages contain smoothly generated counterfactual passages, which could potentially confuse the utility judgments capabilities of the \acp{LLM}. 

\textbf{Listwise}. There are two forms, i.e., listwise-set and listwise-rank in the listwise forms. 
    \begin{enumerate*}[label=(\roman*)]
        \item In the listwise-set form, The utility judgments of \acp{LLM}, particularly ChatGPT, excels on the NQ dataset compared to HotpotQA and MSMARCO-QA datasets under listwise-set form. The reason may be that the NQ dataset comprises relatively simple factual questions, where \acp{LLM} demonstrate a superior ability to generate answers, positively influencing utility judgments.    
        \item In the listwise-rank form, all LLMs excel on HotpotQA compared to other datasets in ranking scores, possibly due to multiple pieces of ground-truth evidence per question, increasing the probability of ground-truth evidence ranking higher.
    \end{enumerate*}
    
    \textbf{Overall analysis}. 
    \begin{enumerate*}[label=(\roman*)]
    
        \item ChatGPT outperforms other \acp{LLM}, highlighting the challenges of open-source models in zero-shot utility judgments. 
        
         \item For LLMs of the same family, utility judgments generally improve as the scale increases. 
         For instance, Vicuna-13B achieves a 73.92\% F1 improvement over Vicuna-7B when using listwise-set input on the NQ dataset.
         
         \item Except for Vicuna-13B and ChatGPT, \acp{LLM} exhibit superior utility judgments in pairwise form compared to listwise form. \acp{LLM} demonstrate better utility judgments in listwise form than in pointwise form.
    \end{enumerate*}

\heading{The sequence of inputs between the question and passages has important effects on utility judgments} 
In this analysis, we use optimal prompts from Fig.~\ref{fig:prompt} with different sequences of input between the question and passages. 
Fig.~\ref{fig:question_passages_order} shows how the sequence of questions and passages affects ChatGPT's utility judgments, with differing effects seen across datasets and input forms. 
For instance, in the listwise-rank input, ChatGPT prioritizes passages first in the NQ dataset but favors questions first in the MSMARCO-QA dataset.
In the NQ dataset, ChatGPT prioritizes questions first when using the listwise-set form, but favors passages first with the listwise-rank form. To ensure consistent prompt design, our future experiments will adopt the question-first input sequence, as shown in Fig.~\ref{fig:prompt}.

\heading{\acp{LLM} demonstrate sensitivity to the order of ground-truth evidence in the listwise input} 
For the experimental setting in this analysis, 
\begin{enumerate*}[label=(\roman*)]
    \item We directly use the optimal prompt in Fig.~\ref{fig:prompt}.
    \item The position of ground-truth evidence is fixed in the input passage list. 
\end{enumerate*}
Prior research has demonstrated a propensity in retrieval-augmented \acp{LLM} \cite{ren2023investigating}  to prioritize evidence presented in the top position \cite{ pezeshkpour2023large}, and has highlighted the order sensitivity in LLMs \cite{lu2021fantastically}.
Therefore, as shown in Fig.~\ref{fig:positions}, to analyze whether \acp{LLM} exhibit sensitivity to order in utility judgments, we position the ground-truth evidence at different positions under listwise-set and listwise-rank inputs. 
We observe that: 
\begin{enumerate*}[label=(\roman*)]

    \item All \acp{LLM} exhibit a notable sensitivity to the position of ground-truth evidence, showcasing significant fluctuations across different positions. 
    
    \item For the listwise-set form, different \acp{LLM} have different sensitivity to ground-truth evidence positions in the input list. 
    The performance of utility judgments for ChatGPT on three datasets first decreases and then increases as the ground-truth evidence position is lower in the input list.
    For LlaMa2-13B, the performance of utility judgments  first increases and then decreases as the ground-truth evidence position is lower in the input list on three datasets. 

   \item For the listwise-rank input form, \acp{LLM} of the same family with different scales have very different performances in utility judgments capabilities under the listwise-rank form, except for ChatGPT.
   Vicuna-13B shows a gradual decline in utility judgment performance as ground-truth evidence is positioned further towards the end. 
   However, Vicuna-7B performs well only at the specific ground-truth evidence position, displaying almost zero performance at other positions. 
   This indicates that models with smaller parameter scales have significantly poor utility judgment capabilities in the listwise-rank form.
   
    \item The sensitivity of a given \ac{LLM} to the position of ground-truth evidence in the input list can vary across different forms. 
    For instance, Vicuna-13B excels with evidence in the middle for listwise-set input but performs better with evidence at the beginning for listwise-rank input.
 
\end{enumerate*}

In practical retrieval augmentation, information about ground-truth evidence positions is often lacking, and the input may not even contain such evidence. 
Addressing the sensitivity of \acp{LLM} to ground-truth evidence positions is crucial and requires immediate attention. 
To mitigate this, we propose a simple $k$-sampling approach in Section \ref{exp-3}, i.e., shuffling the input passages list multiple times and performing utility judgments task. 
This aims to reduce \acp{LLM}' reliance on specific ground-truth evidence positions.

\begin{figure*}[t]
    \centering
    \includegraphics[width=\linewidth]{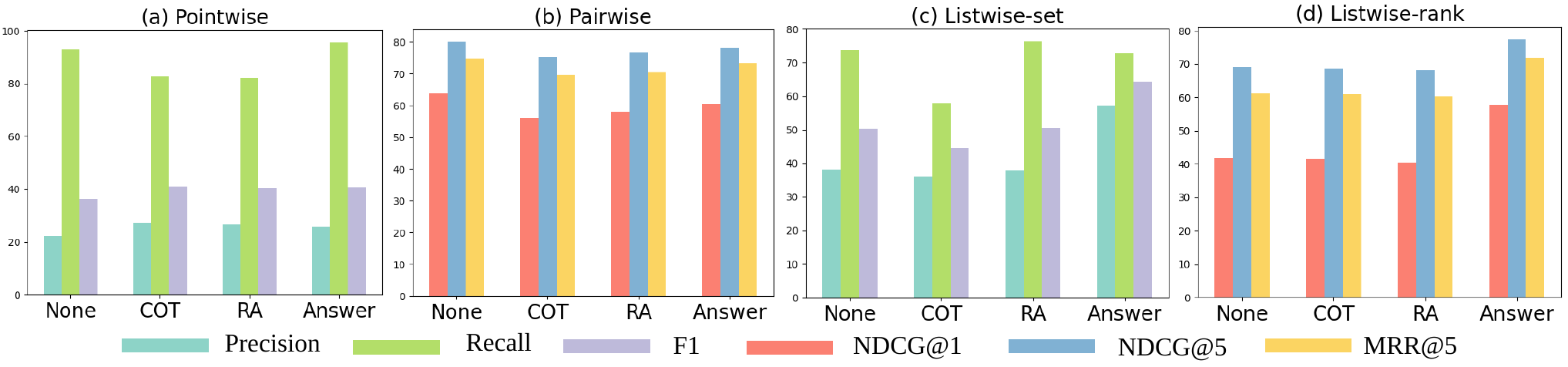}
    \caption{The performance (\%) of utility judgments using ChatGPT under different forms, i.e., pointwise, pairwise, listwise-rank and listwise-set, on the NQ dataset using the prompts with different requirements.}
    \label{fig:prompt_ana}
\end{figure*}

\heading{Utility judgments also depend on additional requirements} 
For the experimental setting in this analysis, 
\begin{enumerate*}[label=(\roman*)]
    \item We directly use the optimal prompt in Fig.~\ref{fig:prompt}.
    
    \item The position of ground-truth evidence is random in the input passage list. 
    
    \item Due to the excessively large number of pairwise input instances, in order to reduce the frequency of API calls, we randomly selected 200 questions from the NQ dataset for pairwise testing. 
    
\end{enumerate*}
We consider three additional requirements in the instructions, i.e., 
\begin{enumerate*}[label=(\roman*)]
    \item \textbf{Chain-of-Thought (COT)} \cite{wei2022chain, kojima2022large} has been proven to be useful for \acp{LLM} in handling complex problems. We incorporate guidance from Zero-shot-CoT \cite{kojima2022large}, i.e., simply adding \emph{``Let's think step by step''} before giving utility judgments. 
    
    \item \textbf{Reasoning (RA)}: Inspired by COT, many NLP tasks have empirically demonstrated performance improvements in LLMs when reasoning is incorporated into prompts \cite{li2023self}. We also design reasoning requirements in prompts like \emph{provide a brief reasoning} before giving the output.
    
    \item \textbf{Answer}: We further guide \acp{LLM} in confirming their utility judgments by having it \emph{``provide the answer to the question''} to the question before giving the output.
    
\end{enumerate*}

Fig.~\ref{fig:prompt_ana} shows the performance of utility judgments using ChatGPT on the NQ dataset. 
We observe that:  
\begin{enumerate*}[label=(\roman*)]
    \item For pointwise form, incorporating COT, RA, and answer requirements enhances F1 performance compared to scenarios lacking additional requirement, possibly due to the challenge posed by limited input information, especially with only one passage available for \acp{LLM} to assess utility directly.

    \item However, when applied to listwise-set inputs with reasoning and listwise-rank inputs with COT, there is a 11.80\% decrease in F1 and a 3.01\% reduction in NDCG@1 compared to no requirement, respectively. 
    This is likely due to the potential influence of noise or incorrect information in passages of varying quality, impacting the reasoning process and overall judgment capability.

    \item Incorporating answer requirements significantly boosts ChatGPT's ability to judge utility in all input forms by implicitly defining passage utility through provided answers.

    \item However, for pairwise inputs, the requirements do not help ChatGPT. 
    E.g., after using the reasoning requirement, the performance of ChatGPT is decreased by 9.37\% in terms of NDCG@1. 
    The reason might be that ChatGPT already has strong pairwise preference judgment capabilities, as evidenced in previous work \cite{jiang2023llm}. 
\end{enumerate*}
  

\vspace{-2mm}
\section{Retrieval-augmented \acp{LLM}: Using selected evidence for QA}\label{exp-3}

The open-domain QA task with \acp{LLM} concerns the process of retrieving external knowledge as evidence and subsequently using the \acp{LLM} to answer questions based on the evidence. 
How do the utility judgments impact the QA abilities of retrieval-augmented \acp{LLM} (\textit{RQ3})?
Specifically, the procedure begins with the retrieval of passages by dense retrievers, without certainty regarding the presence of ground-truth evidence (the GTU setting). 
Then, the utility of the retrieved passages is evaluated by \acp{LLM}, and finally, \acp{LLM} are guided to provide answers based on the selected passages.  

\heading{Experimental setup} 
We evaluate ChatGPT and Vicuna-13B on the NQ and MSMARCO-QA datasets incorporated into the GTU benchmark. 
We employ various types of knowledge as evidence for answer generation: 
\begin{enumerate*} [label=(\roman*)]
    \item No evidence input (\textbf{None});
    
    \item Directly using the top-10 retrieval results as evidence, i.e., candidate passages in the GTU benchmark (\textbf{Dense});
    
    \item Ground-truth evidence (\textbf{Ground-truth});
    
    \item Passages with \textbf{relevance}: using GTU's candidate passages for \acp{LLM} in relevance judgments using the optimal prompt from Fig.~\ref{fig:prompt} under listwise forms; and

    \item Passages with \textbf{utility}: using GTU's candidate passages for \acp{LLM} in utility judgments using the optimal prompt from Fig.~\ref{fig:prompt} under all forms.
\end{enumerate*}
To ensure fairness, the number of results obtained in the form of sets and in the form of a ranked list, after selecting candidate passages, is the same when used in answer generation.

\begin{table*}[t]
  \centering
  \caption{Performance (\%) of question answering using different evidence and different \acp{LLM}. Bold indicates the best answer generation performance among different methods for evidence other than using ground-truth evidence.}
   \renewcommand{\arraystretch}{0.95}
   \setlength\tabcolsep{2.3pt}
    \begin{tabular}{l  cc ccccc cc ccccc }
    \toprule
   \multicolumn{1}{l}{\multirow{3}[6]{*}{Evidence}} & \multicolumn{7}{c}{ChatGPT} &  \multicolumn{7}{c}{Vicuna-13B} \\
     \cmidrule(r){2-8} \cmidrule(r){9-15}   &  \multicolumn{2}{c}{NQ} &  \multicolumn{5}{c}{MSMARCO-QA} &  \multicolumn{2}{c}{NQ} &  \multicolumn{5}{c}{MSMARCO-QA}   \\ 
     \cmidrule(r){2-3} \cmidrule(r){4-8} \cmidrule(r){9-10}\cmidrule(r){11-15}
  & EM  & F1   & ROUGE-L & BLEU-1 & BLEU-2 & BLEU-3 & BLEU-4  & EM  & F1   & ROUGE-L & BLEU-1 & BLEU-2 & BLEU-3 & BLEU-4   \\
  \midrule
   None & 42.49  & 54.55  & 29.78 & 22.64  & 13.63 & \phantom{1}9.10 &  \phantom{1}6.41  & 12.40  & 25.09  & 27.41 & 18.34 & 10.82  & \phantom{1}7.09  & \phantom{1}4.91 \\ 
  Dense &  46.54 &  57.00 & 35.07 & 25.58 & 17.48 & 13.15 &  10.39   & 21.52 & 36.84  & 29.69 & 17.14 & 11.83 & \phantom{1}8.93 & \phantom{1}7.11  \\ 
  Ground-truth & 66.40 & 76.86 & 51.07 & 40.78 & 33.46  & 28.52 & 24.73 & 34.73 & 52.19 & 48.95 & 36.00 & 29.72 & 25.47 & 22.25 \\
  \midrule
 \multicolumn{15}{c}{Relevance judgments}  \\
  \midrule
  Listwise-set  & 47.29 & 57.30 &  35.11 & 25.66 & 17.46  & 13.07 & 10.30 &  21.47 & 36.26 & 30.55 & 18.49 & 12.70 & \phantom{1}9.56 & \phantom{1}7.58  \\
  Listwise-rank   & 47.07 & 57.14  & 35.41 & 25.81 & 17.65 & 13.27 & 10.46 & 21.20 & 36.37 & 30.45 & 18.38 & 12.60 & \phantom{1}9.48 & \phantom{1}7.51 \\
  \midrule
  \multicolumn{15}{c}{Utility judgments}  \\
  \midrule
  Pointwise & 46.16 & 56.59 & 34.51 & 25.41 & 17.18 & 12.84 & 10.12 & 20.61  & 36.58 & 30.26  & 17.82 & 12.33 & \phantom{1}9.34 & \phantom{1}7.46 \\
  
  Pairwise  & \textbf{49.97}  & \textbf{62.06} & 34.86 & 25.82 & 17.37 & 12.90 & 10.08 & 23.24  & 38.31 & 30.98 & 19.64 & 13.35 & 10.00 & \phantom{1}7.91\\
  
  Listwise-set   & 47.72 & 58.01 & 35.68 & 26.52 & 18.15 & 13.68 & 10.85  &  24.10 &  39.07 & 31.00 & 19.04 & 12.92 & \phantom{1}9.68 & \phantom{1}7.69 \\
  
  Listwise-rank  & 48.63  & 58.76 & 35.62 & 26.55 & 18.12 & 13.66 & 10.84 &  23.40 & 37.86 & 30.71 & 19.68 & 13.29 & \phantom{1}9.94 & \phantom{1}7.86 \\
  
  5-sampling & 48.90 & 58.97 & 35.97 & 26.83 & 18.31 & 13.78 & 10.90 & 24.91 &  40.10 & 31.28 & 19.30 & 13.15 & \phantom{1}9.86 & \phantom{1}7.81 \\
  
  10-sampling & 49.49 &  59.66   & \textbf{36.00} & \textbf{26.85} & \textbf{18.33} & \textbf{13.81} & \textbf{10.94} & \textbf{25.39}  & \textbf{40.56} & \textbf{31.59} & \textbf{19.87} & \textbf{13.50} & \textbf{10.11} & \textbf{\phantom{1}8.00} \\

    \bottomrule
    \end{tabular}%
  \label{tab:results-3}%
\end{table*}%

\heading{Different sources of evidence improve the performance of answer generation to different extents} 
From Table \ref{tab:results-3}, we can observe that 
\begin{enumerate*}[label=(\roman*)]

    \item  Using external evidence from a dense retriever markedly improves \acp{LLM}'s answer generation compared to not using external evidence, emphasizing the crucial role of retrieval enhancement in open-domain QA tasks.

    \item  Both utility judgments and relevance judgments enhance \acp{LLM}'s answer generation performance, demonstrating that employing either relevance or utility can effectively filter input passages as evidence.

    \item Using utility judgments for evidence yields better performance in enhancing answer generation compared to using evidence based on relevance judgments in the same input form, further reflecting that \acp{LLM} can distinguish relevance and utility, and then select evidence that has more utility in answering questions.

    \item The listwise-set input outperforms other input forms in answer generation on the MSMARCO-QA dataset using both \acp{LLM}. Meanwhile, the pairwise input achieves the highest performance on the NQ dataset using ChatGPT among all utility judgments approaches, but its real-life implementation involves prohibitively high input costs.
    \item Overall, the degree to which evidence from different sources improves the performance of answer generation is as follows: \textit{Ground-truth} $>$ \textit{Utility judgments} $>$ \textit{Relevance judgments}$>$ \textit{Dense}  $>$ \textit{None}.
\end{enumerate*}

\heading{Novel $k$-sampling listwise approach} According to the conclusions in Section \ref{exp-2}, \acp{LLM} exhibit sensitivity to the position of ground-truth evidence in listwise inputs when judging utility. We propose a $k$-sampling listwise approach (we only use the listwise-set input form as an example). Specifically, we randomize the input passage list $k$ times, conduct utility judgments for each iteration, and then aggregate the results through voting. The evidence chosen for answer generation is determined by the highest vote count. For each query, the number of evidence for answer generation is based on the most frequently occurring listwise-set result across the $k$ iterations.
From Table \ref{tab:results-3}, we can observe that the performance using the $k$-sampling method, such as 10-sampling, improves answer generation on the NQ dataset by 2.84\% in terms of F1 compared to not using sampling in the listwise-set input form. 
Moreover, the $k$-sampling method demonstrated superior answer generation performance, surpassing the usage of other evidence except ground-truth evidence.
The performance improvement indicates that the use of k-sampling effectively mitigates the \acp{LLM}'s dependence on the position of ground-truth evidence. 


\vspace{-2mm}
\section{Related Work} \label{related_work}

\textbf{\acp{LLM} for relevance judgments.} 
With exhibited unprecedented proficiency in language understanding, large language models (\acp{LLM}) such as ChatGPT \cite{chatgpt} and Llama 2 \cite{touvron2023llama} have seen widespread applications across various tasks \cite{izacard2022few, hou2023large, ni2024llms, wang2022cort}. 
IR is a representative work of \acp{LLM} applications, with many studies incorporating \acp{LLM} into relevance ranking  \cite{qin2023large, zhuang2023setwise, liu2023topic, liu2023black}. 
Research into \acp{LLM} in relevance ranking mainly contains the following three approaches: 
\begin{enumerate*}[label=(\roman*)]
\item pointwise \cite{zhuang2023beyond, nogueira2019passage},
\item pairwise \cite{qin2023large, jiang2023llm}, and
\item listwise \cite{sun2023chatgpt, pradeep2023rankvicuna, zhuang2023setwise}.
\end{enumerate*}
\citet{zhuang2023beyond} employed \acp{LLM} in scoring fine-grained pointwise relevance labels.
\citet{qin2023large} employed a pairwise relevance comparison method to distinguish differences between candidate outputs. 
Previous works \cite{sun2023chatgpt, pradeep2023rankvicuna, zhuang2023setwise} analyzed the capabilities of \acp{LLM} in the relevance ranking task.

\citet{faggioli2023perspectives} demonstrated \acp{LLM}' proficiency in relevance assessment in IR. However, relevance in IR and utility in answering specific questions are distinct concepts. This paper investigates whether \acp{LLM} excel in judging the utility of retrieved passages. Similar to relevance ranking tasks, we devise pointwise, pairwise, and listwise approaches for utility judgments.

\heading{Retrieval-augmented \acp{LLM} for QA}
The application of \acp{LLM} in QA \cite{ren2023investigating, yu2022generate, izacard2022few, shi2023replug} is mainly retrieval-augmented \acp{LLM} \cite{ram2023context, ren2023investigating, zamani2022retrieval, fan2024right, fan2024reformatted}. 
Current researches on retrieval-augmented \acp{LLM} can be categorized into two main groups, i.e., independent architectures \cite{xie2023adaptive, mallen2023not, yu2022generate} and joint architectures \cite{izacard2022few, shi2023replug, zhang2023relevance, liu2023webglm}.

In independent architectures, the retriever and \acp{LLM} operate independently, with the retriever's sole role being to provide relevant external knowledge to the \acp{LLM} \cite{zhang2023relevance}. 
For example, \citet{yu2022generate} demonstrated that using retrieval-augmented methods can improve GPT-3 performance on open-domain question answering. 
However, these retrieval models are usually based on the probability ranking principle (PRP) \cite{zhang2023relevance}, ranking passages based on their likelihood of being relevant to the question \cite{zamani2022retrieval, zhang2023relevance}, which may not align with a retrieval-augmented framework. 
In the joint architecture, the \acp{LLM} actively engage in the training process of the retriever \cite{izacard2022few, shi2023replug, lewis2020retrieval, zhang2023relevance}. 
\citet{shi2023replug} used the performance of the \acp{LLM} in answer generation as feedback to train the retriever to retrieve the evidence that contribute more utility to answering the question. 

The independent retriever may struggle to align well with the utility requirements of \acp{LLM} on the retrieval passages. 
Although joint architecture partially alleviates this issue, depending on the answers outputted by \acp{LLM} as utility judgments for retrieved passages is influenced by the \acp{LLM}' internal knowledge. 
Since \acp{LLM} may produce different answers for the same input passages, assessing passage utility based solely on the quality of \acp{LLM}' answers in joint architecture may not always accurately reflect the passages' inherent utility for answering questions.
Therefore, we directly investigate the \acp{LLM}'s capability of utility judgment. 
We hope our work provides useful insights for understanding and improving retrieval-augmented \acp{LLM} in the future.

\section{Conclusion} \label{conclusion}
We have studied the abilities of \acp{LLM} to produce utility judgments for passages.
We have found that \acp{LLM} have different understandings of utility and relevance. 
Moreover, we have shown that utility judgments of \acp{LLM} are influenced by the input forms and positions of ground-truth evidence in the input list, none of which may be a desired property for retrieval-augmented \acp{LLM}.
The susceptibility of \acp{LLM} to these external factors could stem from a limited instruction-following capability. 
We anticipate that as \acp{LLM} continue to advance, the influence of these factors on their capabilities will gradually diminish. 
Finally, we have found that using utility judgments can further improve the performance of answer generation compared to relevance judgments. 

As a preliminary exploration into utility judgments within \acp{LLM}, our analysis has solely focused on evaluating the utility of a small set of candidate passages. 
In the future, it is imperative to devise methodologies for assessing the utility of large-scale candidate passages within the \acp{LLM}. 
This is essential for enhancing utility judgments capabilities in practical applications of retrieval-augmented \acp{LLM}. 
Furthermore, we have only scratched the surface in exploring the zero-shot utility judgments of \acp{LLM}. 
It is crucial to investigate additional scenarios, e.g., the few-shot scenario, to further uncover the capabilities of \acp{LLM} in utility judgments.
We hope our work provides a solid evaluation testbed and meaningful insights for understanding, improving, and deploying utility judgments by \acp{LLM} in the future.
We envision a future where an increasing number of research endeavors contribute to the field of utility judgments in \acp{LLM}.

\begin{acks}
This work was funded by the Strategic Priority Research Program of the CAS under Grants No. XDB0680102, 
the National Key Research and Development Program of China under Grants No. 2023YFA1011\\602 and 2021QY1701,  
the National Natural Science Foundation of China (NSFC) under Grants No. 62372431, 
the Youth Innovation Promotion Association CAS under Grants No. 2021100, 
the Lenovo-CAS Joint Lab Youth Scientist Project, 
and the project under Grants No. JCKY2022130C039.  
This work was also (partially) funded by the Dutch Research Council (NWO), under project numbers 024.004.022, NWA.1389.20.\-183, and KICH3.LTP.20.006, and the European Union's Horizon Europe program under grant agreement No.\ 101070212.
All content represents the opinion of the authors,
which is not necessarily shared or endorsed by their respective employers and/or sponsors. 
\end{acks}

\clearpage
\bibliographystyle{ACM-Reference-Format}
\balance
\bibliography{references}

\end{document}